\documentclass[aps, prd, twocolumn, superscriptaddress]{revtex4}
\usepackage{graphicx}
\usepackage{dcolumn}
\usepackage{amssymb}

\begin{document}

\newcommand{\PHI}{\phi}
\newcommand{\vect}[1]{\mathbf{#1}}
\newcommand{\FT}[1]{\tilde{#1}}
\newcommand{\dash}{^{\prime}}
\newcommand{\ddash}{^{\prime\prime}}
\newcommand{\conj}{^*}
\newcommand{\VEV}{\PHI_{0}}
\newcommand{\diff}{\mathrm{d}}
\newcommand{\Sr}{^{\mathrm{S}}}
\newcommand{\Vr}[1]{\!\!\stackrel{\scriptstyle{\mathrm{V}}}{_{\!\!#1}}}
\newcommand{\Tr}[1]{\!\!\stackrel{\scriptstyle{\mathrm{T}}}{_{\!#1}}}

\newcommand{\half}{\frac{1}{2}}
\newcommand{\ScS}{\scriptstyle}
\newcommand{\ScScS}{\scriptscriptstyle}
\newcommand{\tplus}{\tau\ScScS{+}\ScS{1}}
\newcommand{\tminus}{\tau\ScScS{-}\ScS{1}}
\newcommand{\tplushalf}{\tau\ScScS{+}\!\half}
\newcommand{\tminushalf}{\tau\ScScS{-}\!\half}
\newcommand{\xplus}[1]{\vect{x}\!\ScScS{+}\!\ScS\vect{#1}}
\newcommand{\xminus}[1]{\vect{x}\!\ScScS{-}\!\ScS\vect{#1}}
\newcommand{\xplushalf}[1]{\vect{x}\!\ScScS{+}\!\half\ScS\vect{#1}}
\newcommand{\xminushalf}[1]{\vect{x}\!\ScScS{-}\!\half\ScS\vect{#1}}
\newcommand{\xplushalfplus}[2]{\xplushalf{#1}\ScScS{+}\ScS\vect{#2}}
\newcommand{\xplushalfminus}[2]{\xminushalf{#1}\ScScS{-}\ScS\vect{#2}}
\newcommand{\xplushalfplushalf}[2]{\xplushalf{#1}\ScScS{+}\!\half\ScS\vect{#2}}

\newcommand{\Ob}{\ensuremath{\Omega _{\mathrm b}}}
\newcommand{\Obhh}{\ensuremath{\Ob h^{2}}}
\newcommand{\Ol}{\ensuremath{\Omega _{\Lambda}}}

\newcommand{\wire}{-wire}  


\title{CMB power spectrum contribution from cosmic strings using
  field-evolution simulations of the Abelian Higgs model}

\newcommand{\addressSussex}{Department of Physics \&
Astronomy, University of Sussex, Brighton, BN1 9QH, United Kingdom}

\author{Neil Bevis} 
\email{n.a.bevis@sussex.ac.uk}
\affiliation{\addressSussex}

\author{Mark Hindmarsh} 
\email{m.b.hindmarsh@sussex.ac.uk}
\affiliation{\addressSussex}

\author{Martin Kunz}
\email{martin.kunz@physics.unige.ch}
\affiliation{\addressSussex}
\affiliation{D\'epartement de Physique Th\'eorique, Universit\'e de Gen\`eve, 1211 Gen\`eve 4, Switzerland}

\author{Jon Urrestilla}
\email{jon@cosmos.phy.tufts.edu}
\affiliation{\addressSussex}
\affiliation{Institute of Cosmology, Department of Physics and Astronomy, Tufts
University, Medford, MA 02155, USA}
\date{15 February 2007}

\begin{abstract}
We present the first field-theoretic calculations of the contribution made by cosmic strings to the temperature power spectrum of the cosmic microwave background (CMB). Unlike previous work, in which strings were modeled as idealized one-dimensional objects, we evolve the simplest example of an underlying field theory containing local U(1) strings, the Abelian Higgs model. Limitations imposed by finite computational volumes are overcome using the scaling property of string networks and a further extrapolation related to the lessening of the string width in comoving coordinates. The strings and their decay products, which are automatically included in the field theory approach, source metric perturbations via their energy-momentum tensor, the unequal-time correlation functions of which are used as input into the CMB calculation phase. These calculations involve the use of a modified version of CMBEASY, with results provided over the full range of relevant scales. We find that the string tension $\mu$ required to normalize to the WMAP 3-year data at multipole $\ell = 10$ is $G\mu = [2.04\pm0.06\textrm{(stat.)}\pm0.12\textrm{(sys.)}] \times 10^{-6}$, where we have quoted statistical and systematic errors separately, and $G$ is Newton's constant.  This is a factor 2-3 higher than values in current circulation. 
\end{abstract}


\maketitle


\section{Introduction}
\label{sec:intro}

Observations of the cosmic microwave background (CMB) radiation have helped establish a strong case for cosmic structure to have grown from primordial perturbations created via inflation. On the other hand, theories in which the seeding of structure is primarily attributed to the presence of topological defects \cite{Vilenkin:1994book} have failed to match the data. These defects, of which cosmic strings \cite{Hindmarsh:1994re} are the prime example, have therefore been relegated to at most secondary phenomena with CMB measurements providing upper limits on their relative importance. Although the precise constraints given by current data depend upon the defect model and the data sets chosen, recent calculations have shown that their maximum allowed effect upon the CMB temperature anisotropies is at a level of around 10\% in the power spectrum \cite{Bevis:2004wk, Wyman:2005tu, Fraisse:2006xc}. 

Nevertheless there is once again great interest in topological defects, particularly in the case of cosmic strings, thanks to recent theoretical and observational results (see \cite{Davis:2005dd} for a review). Cosmic strings seem to be viable post-inflation entities in supersymmetric grand unified theories (GUTs) \cite{Jeannerot:2003qv}, superstring theory \cite{Sarangi:2002yt, Copeland:2003bj, Dvali:2003zj, Majumdar:2002hy, Jones:2003da} and hybrid inflation scenarios \cite{Yokoyama:1989pa, Kofman:1995fi}. They have also featured as a possible explanation for the lens candidate CSL-1 \cite{Sazhin:2003cp,Sazhin:2004fv, Sazhin:2005fd, Fairbairn:2005zs}, which initially appeared to be two images of the same galaxy, without the arc-like distortions that would result from lensing by a spheroidal distribution of matter. While new HST data \cite{Sazhin:2006fe} has revealed that CSL-1 is actually two interacting galaxies, this story now serves to highlight one means of cosmic string detection. In a second observational case, an oscillating loop of cosmic string has been discussed as a possible explanation for the synchronous brightness fluctuations in the two images of the (normal) gravitational lens system Q0957+561 \cite{Schild:2004uv}. 

In the present work we consider the effects of an entire network of cosmic strings on the CMB, with the hope that they may be detected by future measurements. Such observations would allow inferences to be made about the network properties, which in principle depend on the nature of the underlying theory and so provide an important window on physics at very high energy. Notably, we present the first CMB calculations for cosmic strings to employ simulations of a field theory, in this case the Abelian Higgs model, which contains \mbox{(local) U(1)} strings. This is a considerable computational challenge that has been made possible via the use of two extrapolations, each justified by the results from the simulations themselves. These enable us to ascertain the contribution to the CMB power spectrum from (traditional) cosmic strings and to compare our results with previous determinations, discussed momentarily, which have involved a greater degree of modelling but have the advantage that they are less computationally demanding.

The strings are considered to have formed at the end of inflation or in a later phase transition. They would then additionally perturb the cosmic fluids at all subsequent times and there would hence be two sources of anisotropy in the CMB. These may in fact be taken to give independent contributions to the CMB power spectrum, and therefore here we calculate that from strings which is to be added to the dominant contribution from inflation. Even if the strings are directly related to the inflation mechanism, their complex non-linear evolution destroys correlations with earlier times. Further, the homogeneity of the CMB implies that the effect of the inflationary perturbations on the strings will be negligible and can be ignored in linear cosmological perturbation theory. Therefore, the perturbations in a quantity $\FT{X}(\vect{k})$ arising from the two mechanisms can be calculated separately and the power spectrum given by:
\begin{equation}
 \left<\! \FT{X}\conj  \FT{X} \!\right> 
 = 
 \left<\! \FT{X}_{\mathrm{inf}}\conj \FT{X}_{\mathrm{inf}} \!\right> 
 +
 \left<\! \FT{X}_{\mathrm{def}}\conj \FT{X}_{\mathrm{def}} \!\right> 
 +
 2 \left<\! \FT{X}_{\mathrm{inf}}\conj \FT{X}_{\mathrm{def}} \!\right>\!,
\end{equation}
but with the cross-term simply zero. The normalization of the string component is then a free parameter related to the energy-scale of the model or equivalently the tension $\mu$, normally expressed as the dimensionless combination $G\mu$. While we present our calculation method and results for the CMB power spectrum here, the precise constraints on $G\mu$ will be addressed in a separate publication \cite{Bevis:2007gh}.


\subsection{Previous cosmic string power spectra}

All previous determinations of the CMB power spectrum for local cosmic strings have relied upon modelling the strings as idealized objects of infinitesimal width. In reality, the string width is related inversely to the energy scale present in the corresponding theory, which could be as high as the grand unification scale $10^{16}$ GeV, whereas the string separation is comparable to the Hubble radius (corresponding to \mbox{$10^{-42}$ GeV} in the present epoch). The idealization is hence justified by the fact that, at times of importance for CMB calculations, the difference in scale is enormous. 

Past results have stemmed from employing this approximation using either \mbox{(i) Nambu-Goto} simulations of connected string segments \cite{Allen:1996wi, Allen:1997ag, Contaldi:1998mx, Landriau:2003xf}; or \mbox{(ii) a model} involving a stochastic ensemble of unconnected segments \cite{Albrecht:1997nt, Pogosian:1999np, Wyman:2005tu, Pogosian:2006hg}. The first case involves simulations of the dynamical equations for ideal relativistic strings, either in a Friedman-Robertson-Walker (FRW) \cite{Allen:1996wi, Allen:1997ag, Landriau:2003xf} or Minkowski \cite{Contaldi:1998mx} space-time. In (ii), the segments are randomly selected for removal so as to give sub-horizon decay, with the particular parameters of the model chosen to provide a match to, for example, the segment density seen in simulations. Although this approach involves a greater degree of modelling, the CMB results match the form given by the more computationally intensive simulations of Contaldi et al. \cite{Contaldi:1998mx}. However, it should be pointed out that the Contaldi et al. \cite{Contaldi:1998mx} (and derived papers \cite{Contaldi:1998qs, Copeland:1999gn}) present the only previously published simulation-based CMB calculations for the very angular scales ($\ell=300-700$) at which the string contribution is shown in to peak in that reference. 

Unfortunately, there have been questions raised as to the accuracy of the Nambu-Goto simulations themselves with regard to small-scale structure, loop production and the decay of strings. For example, in the Minkowski space-time simulations using the Smith-Vilenkin algorithm (as used by Contaldi et al. \cite{Contaldi:1998mx}), the only means via which the network can lose energy is in the removal of small loops, created by self-intersection events. In the conventional picture, these loops would decay into gravitational radiation. However in those simulations, loop production is seen to occur most frequently at the smallest scales involved \cite{Vincent:1996rb}, which is the lattice grid in the Smith-Vilenkin algorithm.  Recent simulations using different algorithms in Minkowski space \cite{Vanchurin:2005yb,Vanchurin:2005pa} or in an FRW universe \cite{Ringeval:2005kr, Martins:2005es} point to the initial correlation length as the dominant scale of loop production, although one group claims some sign of evolution toward a scaling form for the loop production function \cite{Martins:2005es}. 

It may be the case that the true Nambu-Goto physics involves increasing energy density in loops of smaller and smaller size until the Nambu-Goto approximation fails close to the string width. If so, then particle production may be more important than gravitational radiation as the means of string decay. A second possibility is that loop production peaks at some small fraction of the horizon scale and it is simply that this has yet to be observed in even the largest Nambu-Goto simulations to date.

\subsection{Power spectra from field simulations}

In contrast to the above, the present approach to CMB calculations for cosmic strings employs simulations in which the string width itself is resolved. That is, the Abelian Higgs model (in the classical approximation) is represented on a lattice and the strings are evolved in terms of their constituent fields. Although this is the simplest model to exhibit local strings, available computational resources limit the size of the simulation volume to a few hundred times the string width. Given the above horizon size considerations, the epoch of interest for the CMB cannot be simulated directly. 

Fortunately, cosmic strings are believed to evolve toward a scaling regime, in which the network appears statistically the same at all times relative to the causal horizon. This enables the statistical results from a small region to be scaled up to larger volumes at later times, which is precisely that required for CMB calculations. However, it is breached at the radiation-matter transition and hence it is the case that scaling solutions in both eras must be studied. 

In fact, scaling has previously been employed in all high resolution power spectrum calculations ($\ell\!>\!300$), either in unconnected segment models to control string decay or to boost the range of scales over-which Nambu-Goto string simulations can provide data. Such ideal simulations are scale-free, however the incorporation of the string width in the present case means that scaling is used to explicitly translate the results to larger scales and as well as boosting the dynamic range. 
 
\begin{figure*}
\resizebox{\textwidth}{!}{\includegraphics{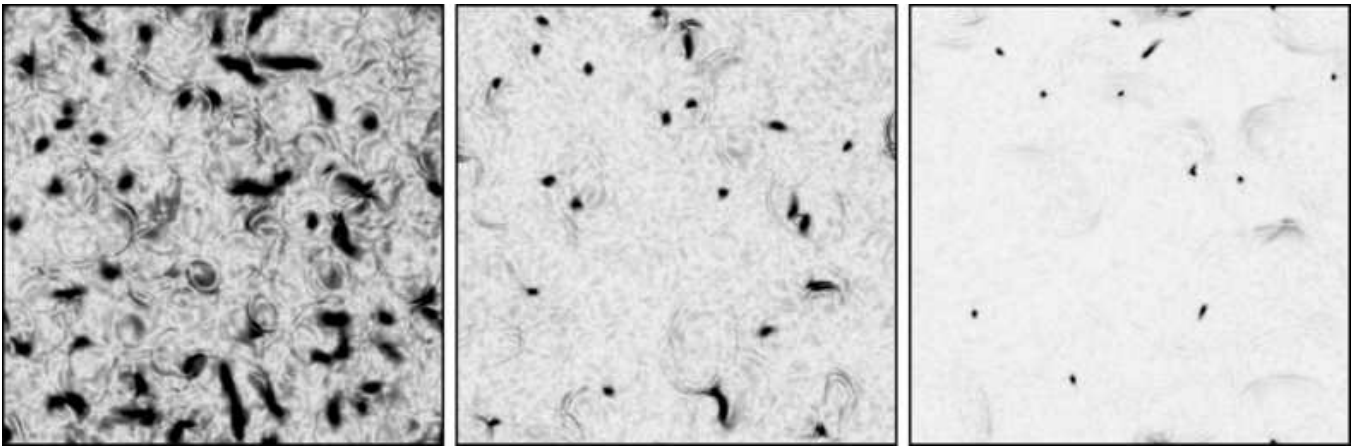}}
\caption{\label{fig:slices}Slices through a $512^{3}$ simulation in the radiation era, showing the Abelian Higgs analogue of magnetic flux density. Magnetic flux tubes run along the cosmic strings, which appear as dark regions, with a shape dependent upon the nature of the string intersection with the slice. Varying left to right, the horizon size (measured as $2\tau$) is 0.35, 0.63 and 1.00 times the box-side while the string width decays inversely with the scale factor $a$ and hence as $\tau^{-1}$ ($a \propto \tau$). Note the decay products visible in these images.}
\end{figure*}

Studies of the Nambu-Goto dynamics and of the field micro-physics provide complementary illuminations of cosmic string dynamics and while field simulations have a smaller dynamic range, they have a number of attractive properties, particularly with regard to CMB calculations. Firstly, the decay products of the strings give rise to additional CMB perturbations but these must be included in an ad hoc manner in Nambu-Goto simulations. Energy is transferred via intersection events from long strings down to small loops, which are traditionally considered to decay into gravitational radiation and are therefore removed from the simulations. It is not the case that gravitational waves are likely to be included self-consistently in any simulation in the near future, due to both the numerical complexity, and technical problems arising from computing the back-reaction of the gravitational radiation on the string network. Contaldi et al. \cite{Contaldi:1998mx} therefore included a compensating fluid via which energy was conserved. On the other hand, particle production is modeled automatically in classical field simulations in terms of oscillations in the fields, which can be seen in \mbox{Fig. \ref{fig:slices}}.

Secondly, field simulations allow a wide class of models to be studied which cannot be approached via the Nambu-Goto approximation. For example, we have already employed the method described here to the additional case of semi-local strings \cite{Urrestilla:2006}, with the present work enabling a comparison between cases without question marks over possible systematics concerning the two simulation approaches. 

Thirdly, a classical field theory simulation is a well-controlled approximation: the 
$\hbar \to 0$ limit of the underlying quantum field theory, and given enough computation even 
the quantum corrections could be computed using stochastic quantization techniques \cite{Berges:2005yt}. The Nambu-Goto approximation is itself an approximation of the classical field theory, and describes only smooth field configurations representing strings with large radius of curvature relative to their width. It fails at cusps and kinks \cite{Blanco-Pillado:1998bv, Olum:1998ag} which are inevitable features of a mobile and intersecting string network.

Field simulations have previously been involved in CMB calculations for the case of global defects \cite{Pen:1997ae, Durrer:1998rw}. In the global case, the defect cores contribute a small fraction of the energy compared to the large-scale field variations and therefore the cores may be left unresolved, with the non-linear $\sigma$-model used to approximate the dynamics. However, in the local string case, the string cores are an important source of energy and momentum and they must be resolved on the computational grid. This presents a serious problem: the string width is a fixed physical length scale, which rapidly decreases in the comoving coordinates used to represent a region of an expanding universe. Either the strings are too narrow to be resolved at the end of the simulation, or so large initially that a representative network cannot be formed. This is illustrated in \mbox{Fig. \ref{fig:slices}}, where it is apparent from the left-most image that, in this large simulation during the radiation era, the string width is not greatly smaller than the string separation. Strings cannot form greatly earlier but the horizon is already around one-third of the box-side. And that is despite the string width being close to the resolution limit at the end of the simulation. In the radiation era the cosmic scale factor $a$ is proportional to conformal time $\tau$, but in the matter era it varies quadratically: \mbox{$a\!\propto\!\tau^{2}$}. This makes the effect even more of a problem in the matter era and the available range in $\tau$ is very limited. 

It is hence required to adopt an approach similar to that used in the Abelian Higgs simulations of Moore et al. \cite{Moore:2001px} (and domain wall simulations of \cite{Press:1989id}), in which the equations of motion were artificially modified such that the string (or domain wall) width became comoving. Focusing on the Abelian Higgs case, Moore et al. have performed simulations in the radiation era, both with and without this artificiality, and report little difference in the dynamics. Unfortunately, the approach introduces a breach of the conservation law for the very energy-momentum tensor through which the strings interact gravitationally with the cosmic fluid. While not especially relevant for the work of Moore et al., this is potentially important for CMB calculations.

The approach used here is therefore to modify the equations so that the comoving string radius $r$ varies with the scale factor as:%
\begin{equation}
 \label{eqn:slimming}
 r 
 \; \propto \; 
 \frac{1}{a^{s}},
\end{equation}
such that $s$ controls the sensitivity of the string width upon $a$. We then make CMB calculations using values of $s$ between $s=0$ (comoving width) and the closest value to the true case ($s=1$) that our facilities permit. Hence in principle, the effect of $s$ upon the CMB results can be ascertained and accounted for.

CMB predictions are made using the unequal-time correlator (UETC) approach \cite{Turok:1996ud, Pen:1997ae}, following the Durrer et al. \cite{Durrer:1998rw, Durrer:2001cg} formalism. In this method, the statistical information taken from the simulations are the two point correlation functions of the energy-momentum tensor: 
\begin{equation}
 \FT{U}_{\lambda\kappa\mu\nu} (\vect{k},\tau,\tau\dash)
 \; = \; 
 \left<
  \FT{T}_{\mu\nu}(\vect{k},\tau) 
  \FT{T}_{\lambda\kappa}\conj(\vect{k},\tau\dash) \;
 \right>
\end{equation}
between unequal times $\tau$ and $\tau\dash$. Although this approach limits the calculation of CMB correlation functions to power spectra, it enables a modified version of an inflation CMB code, CMBEASY \cite{Doran:2003sy} in the present case, to calculate the perturbations created by the defects. As will be explained in Sec. \ref{sec:UETC}, this is via a re-expression of the UETCs as a sum of coherent source functions which drive the string perturbations. Scaling, statistical isotropy, and causality constrain the form of UETCs and their dependence upon the absolute times is trivial. It is merely the ratio $\tau/\tau\dash$ that is important and this enables their application over a large range of times, and therefore CMB scales. This is in contrast to alternative approaches \cite{Pen:1993nx, Allen:1996wi, Landriau:2003xf}, which afford the production of actual CMB maps, useful for non-Gaussianity studies, but are limited to only a small range of scales due to computational constraints. However, the UETC approach still requires the simulations to provide data for the range of time ratios that the UETCs are non-negligible. While UETCs decay for large or small ratios, the range of times over which the strings can be represented is increasingly limited as $s$ is increased. Since the strings take some time after formation to reach scaling to the accuracy required, it is this range of time ratios required that limits the practical value of $s$.


\section{String simulations}
\label{sec:simulations}

\subsection{String model}

The Abelian Higgs model, involving a complex scalar field $\PHI$ and a gauge field $A_{\mu}$, has Lagrangian density:
\begin{equation}
 \label{eqn:lagrangian}
 \mathcal{L}
 = 
 - \frac{1}{4e^{2}} F_{\mu\nu} F^{\mu\nu}
 + (D_{\mu} \PHI)\conj (D^{\mu} \PHI)
 - \frac{\lambda}{4} \left( |\PHI|^{2} - \VEV^{2} \right)^{2}\!\!.
\end{equation}
Here $D_{\mu} = \partial_{\mu} + i A_{\mu}$ is the gauge-covariant derivative and $F_{\mu\nu}=\partial_{\mu} A_{\nu} -
\partial_{\nu} A_{\mu}$ is the field strength tensor, with $e$ and $\lambda$ dimensionless coupling constants. Note that relative to many references, the gauge field is rescaled using $e$, which proves useful when controlling the dependence of the string width upon $a$.

The model obeys the local U(1) symmetry:
\begin{eqnarray}
 \PHI 
 & \rightarrow &
 \PHI \, \exp(i \alpha)
\\
 A_{\mu}
 & \rightarrow &
 A_{\mu} - \partial_{\mu}\alpha,
\end{eqnarray}
which is broken in the low-temperature regime (without quantum corrections to the potential) by any choice of vacuum $|\PHI|=\VEV$. That the vacuum manifold is a closed loop allows for topologically stable string defects, around which the phase of $\PHI$ has a net winding of $2\pi n$ ($n \in \mathbb{Z}^{\pm}$). Although the phase is not a 
gauge-invariant quantity, such a winding would require an infinite gauge transform for its removal. However, the gauge field counteracts the gradient energy associated with the winding such that, in contrast to global strings, the energy of the configuration is localized to a comoving radius: 
\begin{equation}
 \label{eqn:radiusPhi}
 r 
 \; \sim \; 
 \frac{1}{a \sqrt{\lambda} \; \VEV},
\end{equation}
within which $|\PHI|$ approaches zero. As the gauge field attempts to counter the phase gradients near the core, it itself acquires a significant curl around the string, resulting in a magnetic flux tube which traces the string. For the present work we adopt the Bogomol'nyi ratio $\lambda/2 e^{2} = 1$, in which the characteristic scales of the magnetic and scalar energies are equal. It is important to note that despite the analogy with electromagnetism, in the broken-symmetry phase, perturbations about the vacuum reveal:
\begin{equation}
 \mathcal{L} 
 \; = \;
 \ldots 
 + \VEV^{2} A_{\mu} A^{\mu} 
 + \ldots,
\end{equation}
and that the gauge field is massive. This model contains no massless modes and therefore all group velocities are less than the speed of light, except for waves along the strings themselves. Hence, the normal causal run-time limit imposed by the simulation boundaries may be extended as the strings are highly curved and therefore no disturbances can traverse the box by this time.

Variation of the action corresponding to Eq.\ (\ref{eqn:lagrangian}) and making the gauge choice $A_{0}=0$ yields
the dynamical equations for a flat FRW space-time as:
\begin{eqnarray}
 \label{eqn:AHdynamicPhi}
 \ddot{\PHI} + 2 \frac{\dot{a}}{a}\dot{\PHI} - D_{j} D_{j} \PHI 
 & = &
 -a^{2} \frac{\lambda}{2} \left( |\PHI|^{2} - \VEV^{2} \right) \PHI
\\
 \label{eqn:AHdynamicF}
 \dot{F}_{0j} - \partial_{i}F_{ij}
 & = &
 -2a^{2} e^{2} \; \mathcal{I}m \! \left[ \PHI\conj D_{j}\PHI \right].
\end{eqnarray}
The notation here is such that over-dots denote differentiation with respect to conformal time $\tau$, and $\partial_{i}$ with respect to comoving Cartesian coordinates. Further, the system obeys the constraint:
\begin{equation}
 -\partial_{i} F_{0i} = -2 a^{2} e^{2} \; \mathcal{I}m [ \PHI\conj \dot\PHI ],
\end{equation}
which is analogous to Gauss' law, and thereby conserves the 4-current density: $-2a^{2}e \; \mathcal{I}m [ \PHI\conj D_{\mu}\PHI ]$. 

It is the presence of the $a^{2}$ factors on the right-hand-sides of these equations that causes the string width to lessen in comoving coordinates and causes the problem highlighted in the previous section. Moore et al.  \cite{Moore:2001px} have evolved Eq.\ (\ref{eqn:AHdynamicPhi}) and (\ref{eqn:AHdynamicF}) in the radiation era and then additionally with the factors of $a^{2}$ removed from the right-hand-sides. The result was that not only did the string length show scaling but that the scaling forms were very similar, allowing them to use the modified equations to study the matter era also. However, it is the desire here to have a controllable dependence of the width upon $a$, parametrized by the parameter $s$ according to Eq.\ (\ref{eqn:slimming}). One means of achieving this is to raise the dimensional coupling constants to time-dependent variables according to:
\begin{eqnarray}
 \lambda & = & \lambda_{0} \; a^{-2(1-s)}
\\ 
 e & = & e_{0} \; a^{-(1-s)}.
\end{eqnarray}
Variation of the action now yields the artificial equations of motion:
\begin{eqnarray}
 \label{eqn:AHdynamicPhiS}
 \ddot{\PHI} + 2 \frac{\dot{a}}{a} \dot{\PHI} - D_{j} D_{j} \PHI 
  = 
 -a^{2s} \frac{\lambda_{0}}{2} \left( |\PHI|^{2} - \VEV^{2} \right)\!\PHI \;
\\
 \label{eqn:AHdynamicFS}
 \dot{F}_{\!0j} + 2 (1\!-\!s) \frac{\dot{a}}{a} F_{\!0j} - \partial_{i}F_{\!ij}
  =
 -2a^{2s} e_{0}^{2} \; \mathcal{I}m \! \left[ \PHI\conj D_{j}\PHI \right],
\end{eqnarray}
and the constraint:
\begin{equation}
 -\partial_{i} F_{0i} 
 = 
 -2 a^{2s} e_{0}^{2} \; \mathcal{I}m \! [ \PHI\conj \dot\PHI ].
\end{equation}
Notice that there is an additional damping term for $s<1$ that is created by the lowering of $e$ with time, and is required for the dynamical equations to preserve Gauss' law in this consistent form. Therefore this approach, even in the case of $s=0$, is not precisely the same as that of Moore et al. \cite{Moore:2001px}.


\subsection{Lattice discretization}
It has become standard in the literature \cite{Vincent:1997cx,Moore:2001px} to discretize the Abelian Higgs
model on a lattice via the method of Moriarty et al. \cite{Moriarty:1988fx}. Rather than discretizing
the dynamical equations directly, which does not generally lead to an algorithm that preserves the Gauss constraint, the Moriarty et al. approach is to discretize the Hamiltonian. Then discrete equations of motion, including a discrete version of the constraint equation, can be derived from it. This is broadly the approach followed here, except
that the action is discretized rather than the Hamiltonian, preferable given that these simulations are performed for an FRW rather than a Minkowski metric, and additionally the variables $\lambda$ and $e$ are time-discretized.

The discretization preserves the gauge symmetry in the form:
\begin{eqnarray}
 \PHI^{\vect{x},\tau} 
 & \rightarrow & 
 \PHI^{\vect{x},\tau} \exp(i\alpha^{\vect{x},\tau})
\\
 A_{0}^{\vect{x},\tplushalf}
 & \rightarrow &
 A_{0}^{\vect{x},\tplushalf} 
   - \frac{1}{\Delta \tau} 
     \left( 
       \alpha^{\vect{x},\tplus} - \alpha^{\vect{x},\tau} 
     \right)
\\  
 A_{j}^{\xplushalf{j},\tau}
 & \rightarrow &
 A_{j}^{\xplushalf{j},\tau} 
   - \frac{1}{\Delta x} 
     \left( 
      \alpha^{\xplus{j},\tau} - \alpha^{\vect{x},\tau} 
     \right). 
\end{eqnarray}
The notation is such that $A_{j}^{\xplushalf{j},\tau}$ signifies that the spatial components of the gauge field are represented on the links half-way between lattice steps. This creates a consistent centered-derivative transform that is in-line with the geometric interpretation of the gauge field: that it performs a local rotation of the field-coordinates to form the gauge-covariant derivative. Hence, this derivative is:
\begin{equation}
 \left( D_{j}\PHI \right)^{\xplushalf{j}}
 \; = \;
 \frac{1}{\Delta x} 
 \left( 
      \PHI^{\xplus{j}} - \exp(-i \theta_{j}^{\xplushalf{j}} ) \PHI^{\vect{x}}
 \right),
\end{equation}
with:
\begin{equation}
 \theta_{j}^{\xplushalf{j}}
 \; = \;
 \Delta x \; A_{j}^{\xplushalf{j}},
\end{equation}
making the gauge-field a purely angular variable. 

The entity:
\begin{equation}
 \Delta_{ij}^{\xplushalfplushalf{i}{j}}
 \; = \;
 \left( \theta_{j}^{\xplushalfplus{j}{i}} \!-\! \theta_{j}^{\xplushalf{j}} \right)
 \! - \! 
 \left( \theta_{i}^{\xplushalfplus{i}{j}} \!-\! \theta_{i}^{\xplushalf{i}} \right)
\end{equation}
is invariant under the discrete gauge transform and is used to build the $F_{ij}$ magnetic term in the action. However, following Wilson \cite{Wilson:1974sk}, this is performed in a manner that preserves this angular nature and the total action integral is represented as the discrete sum:
\begin{widetext}
 \begin{eqnarray}
  S 
  \; = \;
  \Delta \tau \; (\Delta x)^{3} \sum_{\tau} \sum_{\vect{x}} 
  \!\left(
    -\frac{1}{2 e_{\tau}^{2}\;(\Delta x)^{4}} 
    \sum_{i} \sum_{j} 
    \left[
    1 - \cos( \Delta_{ij}^{\xplushalfplushalf{i}{j},\tau} )
    \right]
    +
    \frac{1}{2e^{2}_{\tplushalf} (\Delta x)^{2}(\Delta \tau)^{2}} 
    \sum_{i} \left( 
         \theta_{i}^{\xplushalf{i},\tplus} - \theta_{i}^{\xplushalf{i},\tau} \right)^{2} 
             \right. &
\\
    \left.
    +\;
    a_{\tplushalf}^{2} 
    \left| 
      \frac{ \PHI^{\vect{x},\tplus}-\PHI^{\vect{x},\tau} }{\Delta \tau} 
    \right|^{2}
    - 
    a^{2}_{\tau} \sum_{i} 
    \left| 
      \frac{\PHI^{\xplus{i},\tau} e^{i\theta_{i}^{\xplushalf{i},\tau}} - \PHI^{\vect{x},\tau}}
           {\Delta x}
    \right|^{2}
    -
    a^{4}_{\tau} \frac{\lambda_{\tau}}{4} 
    \left( |\PHI^{\vect{x},\tau}|^{2} - \VEV^{2} \right)^{2}
   \right)\!. \nonumber   
 \end{eqnarray}
\end{widetext}
Hence the first term tends to $\frac{1}{4e^{2}}F_{ij}F_{ij}$ in the continuum limit since $1\!-\!\cos x \rightarrow \half x^{2}$ as $x \rightarrow 0$.

In the temporal gauge $A_{0}=0$ used for evolution, there is no requirement to treat the $F_{0i}$ electric term in such a manner and the second term is essentially just the time derivative of the gauge field. Note however that this term involves a centered and therefore mid-step derivative and so $e$ has likewise been referenced mid-step. The situation is then very similar for the third term, which (in this gauge) is just the time derivative of the scalar field, with the scale-factor referenced mid-step. The remaining terms then involve the spatial gauge-covariant derivative and the potential.

Variation of this sum with respect to $\PHI^{\vect{x},\tau}$ or $\theta_{i}^{\xplushalf{i},\tau}$, for a given $\vect{x}$ and $\tau$, then yields the evolution equations. The constraint may be derived by including $A_{0}$ (although there is no requirement to do so via the full Wilson method) and then minimizing the action with respect to it. Preservation is assured (to machine precision) in the temporal gauge due to the discrete gauge symmetry and that the constraint results from an independent minimization of the action sum. The dynamical equations in the chosen gauge allow for a leap frog update algorithm, in which the fields are updated using their stored time derivatives, and then the time derivatives at a site are updated using the field values at that site and its
neighbours. There are no off-site references made to the quantity being updated and therefore there is no additional temporary storage required. The evolution hence involves $10$ single-precision floating point numbers per site, or for an $N^{3}$ lattice, $40 \,(N/1024)^{3}$ GB of data updated each timestep.

The resolution of strings at the end of a simulation is assured by setting $a_\mathrm{end} = 1$, $\lambda_{0}=2$, $e_{0}=1$, and then $\Delta x = 0.5 \VEV^{-1}$, which has been shown to yield good string resolution in Minkowski space-time simulations ($a\equiv1$) \cite{Vincent:1997cx}. The choice of $\Delta \tau / \Delta x = 0.2$ gives good covariant energy-momentum conservation in the $s=1$ case (in the radiation era). For example during an $s=1$ phase in $512^{3}$ simulations, the measured decrease in the mean energy density was found to be within $\pm7\%$ of that expected from Hubble damping. For $s=0.3$ in the radiation era, the rate of energy loss is $10-20\%$ greater than the conservation law would predict. As discussed in the previous sub-section, the absence of massless modes allows the simulation to be run longer than the half-box crossing time for light, with $\tau_{\mathrm{end}}$ = $1.25 (N/2)\Delta x$ used here (for $s<1$ cases). However, these late times are used to provide the least important UETC data. A single simulation requires $\sim\!500$ CPU-hours of processor time on the UK National Cosmology Supercomputer \cite{cosmos-website}. Parallel processing is afforded by MPI via a library, written as part of this work, which allows the rapid development of parallel field-evolution simulations using a convenient C++ object-orientated interface \cite{LATfield}.


\subsection{Initial conditions}
\label{sec:IniCon}
An initial field configuration consisting of a scaling network of strings is difficult to achieve directly. Previous independent simulations of Abelian Higgs string networks \cite{Vincent:1997cx,Moore:2001px} have used different initial conditions and have recovered consistent results for the string length density in Minkowski space-time during the scaling regime. Accepting that the precise nature of the initial conditions is unimportant, it is desirable that they give a rapid convergence toward scaling but with minimal complexity in their generation.
  
The initial field configuration must satisfy the discrete form of Gauss' law and the simplest means of achieving this is to set all temporal derivatives and the gauge field to zero across the entire lattice. Then $\PHI$ is set to have modulus $\VEV$ with a random phase assigned to each lattice site. With the initial time set such that $\Delta x$ is of order the causal horizon, this sets up independent phases in each initial horizon volume and is a rough approximation to the results of a phase transition. The system is then evolved according to the discrete equations of motion and Hubble damping relaxes the system into a network of cosmic strings. This approach is in contrast with Moore et al. \cite{Moore:2001px} in which the network relaxation is achieved using a period of diffusive (first-order) evolution, with the second derivatives removed from the equations of motion.

\begin{figure}
\resizebox{\columnwidth}{!}{\includegraphics{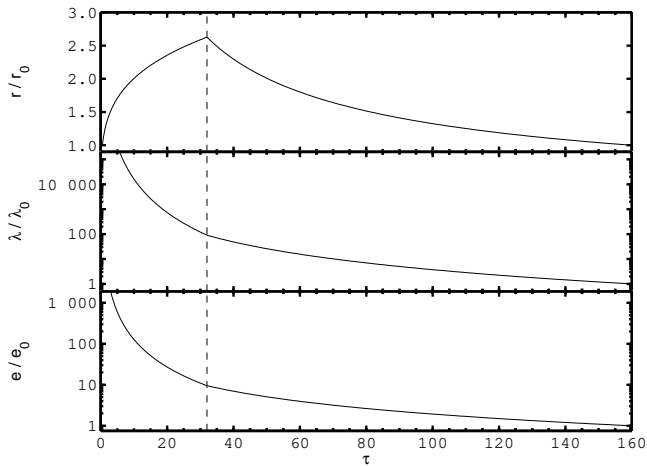}}
\caption{\label{fig:stringWidth}The variation of the string width $r$ and coupling parameters
         with conformal time $\tau$ (in units of $\VEV^{-1}$) for the $s\!=\!0.3$ simulations described in Section  
         \ref{sec:IniCon}. The subscript 0 indicates the value at the end of the simulation.}
\end{figure}

However, strings cannot form until the causal horizon is larger than their characteristic width, since otherwise gradients exist on scales so small that there is sufficient energy for the entire volume to rise up the potential toward $|\PHI|=0$. Although
the choice of $s<1$ lowers the initial string width relative to its final value, strings will still not form for some time. Since the network will take further time to reach the scaling regime, it is desirable to accelerate the formation process so that UETC data may be taken over as large a ratio of times of possible. Hence the parameter $s$ is in fact set to be negative prior to a time $\tau_{s}$ such that the string width in fact \emph{grows} from an initially low value, before $s$ takes on its final positive value and the width shrinks during the second phase of the simulation. The string width hence varies according to Fig.\ \ref{fig:stringWidth} and a network of strings forms by $\tau \approx 10\VEV^{-1}$.

\section{CMB calculation method}
\label{sec:UETC}

\subsection{UETC approach}

The evolution of the cosmological perturbations may be described by a linear differential equation of the following form:
\begin{equation}
 \hat{\mathcal{D}}_{ac}(k,a,\dot{a},\rho,...)
 \FT{X}_{a}(\vect{k},\tau\dash)
 \; = \;
 \FT{S}_{c}(\vect{k},\tau\dash),
\end{equation}
with $\FT{X}_{a}$ denoting the Fourier Transform of $X_{a}$. The linear differential operator $\hat{\mathcal{D}}_{ab}$ includes quantities from the background FRW universe such as the mean physical density $\rho$ and the cosmic scale factor. It acts upon the metric, matter and photon perturbations described by the vector $\FT{X}_{a}$, with the source term $\FT{S}_{c}$ describing the active seeding due to the defect presence --- their energy-momentum tensor. For the homogeneous case corresponding to inflation ($\FT{S}_{c}\!=\!0$), this equation set can be solved using the standard CMBEASY code. In principle therefore, if $\FT{S}_{c}(\vect{k},\tau\dash)$ is known, then this inhomogeneous set can be solved using a Green's function $\mathcal{G}_{ac}(k,\tau,\tau\dash)$ to give the perturbation power spectra for a wave vector $\vect{k}$ and time $\tau$ as:
\begin{equation}
 \label{eqn:greens}
 \left<\! 
   \FT{X}_{a} \FT{X}_{b}\conj
 \!\right>
 =
 \!\int\!\!\!\! \int\!
     \diff\tau\dash \diff\tau\ddash 
     \mathcal{G}_{ac}(\tau\dash\!)
     \mathcal{G}_{bd}\conj(\tau\ddash\!)
     \left<\!
      \FT{S}_{c}(\tau\dash\!) \FT{S}_{d}\conj(\tau\ddash\!)
     \!\right>\!.    
\end{equation}
Although this is not the actual method used here, this equation shows that the data needed from the simulations for CMB power spectra calculations are the bracketted terms on the right. These are the Fourier transforms of the two-point correlation functions:
\begin{equation}
 U_{cd}(\vect{y},\tau, \tau\dash)
 \; = \;
 \frac{1}{V} 
 \int \diff^{3} \vect{x} \;
   \left< S_{c}(\vect{x},\tau) \,
   S_{d}(\vect{x}\!-\!\vect{y},\tau\dash) \right>,
\end{equation}
with the normalization of the Fourier transform chosen as:
\begin{equation}
 \FT{U}_{cd}(\vect{k},\tau,\tau\dash)
 = 
 \frac{1}{V} 
 \int \diff^{3} \vect{y} \;
  U_{cd}(\vect{y},\tau,\tau\dash) \;
  e^{-i \vect{k}\cdot\vect{y}}.
\end{equation}
Here $V$ is the fiducial simulation volume and its inclusion yields $\FT{U}_{cd}$ with the same dimensions as $S_{c}$ (and so $T_{\mu\nu}$) squared: $[\FT{U}_{cd}]\!=\![U_{cd}]\!=\!(\mathrm{time})^{-4}$.  

For scaling sources, a statistical measure of the dynamics should be dependent upon a single scale $d$. While the energy-scale $\VEV$ should not affect the spatial distribution of strings, it does set the normalization of the energy-momentum tensor as $\VEV^{2}$. Hence assuming scaling, $U_{bc}$ may be written as:
\begin{equation}
 U_{cd}(y, \tau, \tau\dash) 
 \; = \;
 \frac{\VEV^{4}}{d^{4}} 
 \;
 f \! \left( 
     \frac{y}{d},\frac{\tau}{d},\frac{\tau\dash}{d}
   \right)\!,
\end{equation} 
with $f$ a dimensionless function. The scale $d$ must be symmetric in the two times involved and hence may be written as:
\begin{equation}
 d
 =
 \sqrt{\tau\,\tau\dash} \, g(\tau/\tau\dash)
 =
 \sqrt{\tau\,\tau\dash} \, g(\tau\dash/\tau).
\end{equation}
In this form the final two inputs in $f$ provide the same information, and further the dimensionless function $g$ can be absorbed to yield:
\begin{equation}
 U_{cd}(y,\tau,\tau\dash) 
 \; = \;
 \frac{\VEV^{4}}{(\tau\,\tau\dash)^{2}} 
 \;
 F\!\left(\!
       \frac{y}{\sqrt{\tau\,\tau\dash}},\frac{\tau}{\tau\dash}
    \!\right)\!.
\end{equation}
The Fourier transform then gives:
\begin{eqnarray}
 \FT{U}_{cd}(k,\tau,\tau)
 & = &
 \frac{\VEV^{4}}{\sqrt{\tau\,\tau\dash}} \frac{1}{V}
 \;
 \FT{C}(k\sqrt{\tau\,\tau\dash},\tau/\tau\dash).
\end{eqnarray}
The change in the power of $(\tau\,\tau\dash)$ comes from a change in integration variable, required to match the dimensionless spatial input to $F$. Note also that $V$, which is not involved in the dynamics of the system, is left aside in the dimensional analysis.

The resultant scaling function $\FT{C}$, which describes all the unknowns with regard to this UETC, has no associated absolute scale and is a function merely of two variables. Further, since the quantities correlated are real (in real-space), then $\FT{U}\conj_{cd}(\vect{k})\!=\!\FT{U}_{cd}(-\vect{k})$ and statistical isotropy implies that the scaling functions are also real. Hence these functions are an efficient means of summarizing the data from the simulation as well as having the function of scale-extrapolation.

It should be noted however, that the power of the UETC approach stems also from (i) strings decay on scales much smaller than the horizon and (ii) there can be no super-horizon correlations (since the strings form at the end of or after inflation). From (i), the string $\FT{T}_{\mu\nu}(\vect{k},\tau)$ is unimportant for $k\tau \!\gg\! 1$ and since the scaling functions involve a product of two such terms they therefore decay for high $k\sqrt{\tau\tau\dash}$. In the opposite limit, (ii) implies that the scaling functions may be expanded as simple power laws for low $k\sqrt{\tau\,\tau\dash}$ \cite{Durrer:2001cg}. Statistical isotropy then forbids odd terms in any such expansion ($\FT{U}(\vect{k},\tau,\tau\dash)\!=\!\FT{U}(-\vect{k},\tau,\tau\dash)$) and the result is that 4 of the 5 scaling functions required here are simply constant for low $k\sqrt{\tau\,\tau\dash}$. In the final case, it is expected to vary as $k^{2}\tau\,\tau\dash$, which can then be extracted to yield the same constancy for super-horizon scales. Furthermore, if the ratio of times $\tau/\tau\dash$ is very large or very small, then the combination of (i) and (ii) implies that for any $k\sqrt{\tau\,\tau\dash}$, one of the factors of $\FT{T}_{\mu\nu}(\vect{k},\tau)$ is insignificant while one comes from the super-horizon regime. The scaling functions must hence decay for extreme time ratios and are most important for near-equal times.

These simple considerations regarding the form of the scaling functions \cite{Durrer:1997ep} mean that a field simulation, only able to provide a limited range in $k\sqrt{\tau\,\tau\dash}$ and $\tau/\tau\dash$, can still provide the data required for accurate results over a large range of scales.


\subsection{Perturbation Equations}

As in the inflationary case, the perturbations evolve according to the linearized Einstein equations, covariant energy and momentum conservation and the Boltzmann equation \cite{Durrer:2001cg}. The key change in the present case is the additional presence of the string energy-momentum tensor $\FT{T}_{\mu\nu}$ in the Einstein equations. In the linear regime appropriate for CMB calculations, the energy-momentum conservation equations may be separated into those involving the perturbations in the cosmic fluids and those involving the $T_{\mu\nu}$ components, since products of these small quantities can be ignored. Hence, as stated in the Introduction, the defect energy-momentum tensor is separately conserved. However, it also does not suffer from the gauge-dependence of the space-time metric and cosmic fluid perturbations. This is as a result of it being identically zero in the homogeneous background relative to which the perturbations are measured. The metric and fluid perturbations are also described here in terms
of gauge-invariant quantities, via the formalism of Durrer et al. \cite{Durrer:1998rw, Durrer:2001cg} to which the reader is referred for greater detail. 

It is convenient to decompose the perturbations into 3-scalars, 3-vectors and 3-tensors according to their transformations under O(3) rotations. These three classes then evolve according to independent equations and are sourced by independent projections from the string energy-momentum tensor. Whereas vector perturbations decay in the standard inflationary scenario and so are not taken into account, they are continuously sourced in the defect case and must be considered. This is the second major difference between the defect and inflationary cases.

As is common, the space-time metric is decomposed (in Fourier space) into scalar (S), vector (V) and tensor (T) parts as:
\begin{eqnarray}
 \FT{g}_{00} 
 & = &
 a^{2} \left( 1 + \FT{A}\Sr \right)
\\
 \FT{g}_{0i} 
 & = &
 a^{2} \left( \hat{k}_{i} \FT{B}\Sr + \FT{B}\Vr{i} \right)
\\
 \FT{g}_{ij} 
 & = &
 - a^{2} 
 \left( 
  1 + \delta_{ij} \FT{H}\Sr_{\mathrm{L}} 
    + \left[ \hat{k}_{i}\hat{k}_{j} - \frac{1}{3} \delta_{ij} \right]\FT{H}\Sr_{\mathrm{T}}
 \right.\\
 & & \left. \;\;\;\;\;\;\;\;\;\;\;\;\;\;\;\;\;\;
    +\; \frac{1}{2} \left[ \hat{k}_{i} \FT{H}\Vr{j} + \hat{k}_{j} \FT{H}\Vr{i} \right]
    + \FT{H}\Tr{ij} 
 \right) \nonumber.
\end{eqnarray}
In the scalar case, the gauge-invariant Bardeen potentials may then be formed as:
\begin{eqnarray}
 \FT{\Phi}\Sr
 & = &
 \FT{H}\Sr_{\mathrm{L}}
 + \frac{1}{3} \FT{H}\Sr_{\mathrm{T}} 
 - \frac{\dot{a}}{a} \frac{\FT{\Sigma}\Sr}{k}
\\
 \FT{\Psi}\Sr 
 & = &
 \FT{A}\Sr 
 - \frac{1}{k} 
 \left( 
    \frac{\dot{a}}{a} \FT{\Sigma}\Sr 
    - \dot{\FT{\Sigma}\Sr} 
 \right),
\end{eqnarray}
where $\FT{\Sigma}\Sr \!=\! k^{-1}\dot{\FT{H}}{\Sr_{T}} - \FT{B}\Sr$. The Einstein equations for these potentials are then simplified if the two (of the four) scalar degrees of freedom in $T_{\mu\nu}$ are projected out as:
\begin{eqnarray}
 \FT{S}\Sr_{\Phi} 
 & = &
 \FT{T}_{00} - 3 \frac{\dot{a}}{a} \frac{i\hat{k}_{m}}{k} \FT{T}_{0m}
\\
 \label{eqn:defPsi}
 \FT{S}\Sr_{\Psi}
 & = &
 -\FT{S}\Sr_{\Phi} - T_{mm} + 3 \hat{k}_{m}\hat{k}_{n} \FT{T}_{mn}.
\end{eqnarray}
The relevant Einstein equations then appear as:
\begin{eqnarray}
  k^{2} \FT{\Phi}\Sr  
  & \! = \!&
  4\pi G \! 
  \left(
   \FT{S}_{\Phi}\Sr - 3a^{2} (\rho\!+\!p) \FT{\Phi}\Sr \! + \!\ldots
  \right)
\\
  k^{2}
  \left(
   \FT{\Phi}\Sr + \FT{\Psi}\Sr
  \right)
  & \! = \!&
  4\pi G \! 
  \left(
   \FT{S}_{\Phi}\Sr + \FT{S}_{\Psi}\Sr + \ldots
  \right).
\end{eqnarray}
The ellipsis denote the additional presence on the right of the matter and photon perturbations, the notation for which shall not be defined here in a desire for brevity. Note that although these equations relate metric perturbations on the left to energy perturbations on the right, a Bardeen potential does appear on the right due the involvement of the metric tensor in raising and lowering the indices of tensor measures of the FRW background, namely the total background density $\rho$ and pressure $p$. While there are actually four scalar degree of freedom in $\FT{T}_{\mu\nu}$, that this tensor obeys covariant energy-momentum conservation provides two scalar equations via which the remaining two are specified.  

In the case of vector modes, it is useful to define:
\begin{equation}
 \FT{S}\Vr{i}
 \; = \;
 \FT{T}_{0i} - \hat{k}_{i} \hat{k}_{m} \FT{T}_{0m},
\end{equation}
which then obeys the vector constraint $k_{i} \FT{S}\Vr{i} \!=\! 0$ and so contains two of the four vector degrees of freedom in $\FT{T}_{\mu\nu}$. This gives a simple form for the corresponding Einstein equation as:
\begin{equation}
 -k^{2} \FT{\Sigma}\Vr{i}
 \; = \;
 16\pi G \left( \FT{S}\Vr{i} + \ldots \right),
\end{equation}
where $\FT{\Sigma}\Vr{i} \!=\! k^{-1} \dot{\FT{H}}\;{\Vr{i}} - \FT{B}\Vr{i}$.
Momentum conservation then implies that the remaining two vector degrees of freedom, which come from $\FT{T}_{ij}$, can be found from $\FT{S}\Vr{i}$ and its time derivative. 

Of the ten degrees of freedom in $\FT{T}_{\mu\nu}$, the remaining two are 3-tensors and can be projected out of the space-space components in an analogous manner to $\FT{H}\;{\Tr{ij}}$. This results in $S\Tr{ij}$, which sources perturbations via:
\begin{equation}
 \ddot{\FT{H}}\Tr{ij} 
 + 2\frac{\dot{a}}{a} \dot{\FT{H}}\Tr{ij}
 + k^{2} \FT{H}_{ij}
 \; = \;
 8\pi G \left( \FT{S}\Tr{ij} + \ldots \right).
\end{equation}
There is no tensor equation that stems from energy-momentum conservation and hence this property cannot be used to apply further constraint in this case.

The result is that the sourcing of perturbations by strings can be described in terms of the variables $\FT{S}\Sr_{\Phi}$, $\FT{S}\Sr_{\Psi}$, \mbox{$\FT{S}\Vr{i}$} and \mbox{$\FT{S}\Tr{ij}$}, containing six degrees of freedom in total. These appear in the Einstein equations, via which the space-time metric is determined, but do not otherwise influence the remainder of the standard CMBEASY evolution routine. That is with the exception that the continual sourcing of vector modes means that vector contributions must be additionally considered, and are evolved here using the Hu and White method \cite{Hu:1997hp}. 


\subsection{UETC scaling functions}
Initially there are 10 independent components in the string energy-momentum tensor, between which there are potentially $\frac{1}{2}10\,(10\!+\!1)\!=\!55$ UETC scaling functions to be considered. However as noted above, the number of scalar and vector degrees of freedom may be reduced via energy-momentum conservation such that there are two degrees of freedom for each scalar, vector and tensor class. Further, statistical isotropy enforces that UETCs between these classes are simply zero. 

For the scalar class, the three independent UETCs are \cite{Durrer:1998rw, Durrer:2001cg}:
\newcommand{\removeLHspace}{\!\!\! \!\!\! \!\!\!}
\begin{eqnarray}
 \removeLHspace
 \left<\! 
  \FT{S}\Sr_{\Phi}(\vect{k},\tau) \,
  \FT{S}_{\Phi}^{\mathrm{S}*}(\vect{k},\tau\dash) 
 \!\right>
 & \!\!\!=\!\!\!&
 \frac{\VEV^{4}}{\sqrt{\tau\,\tau\dash}} \frac{1}{V}
 \FT{C}\Sr_{11}(k\sqrt{\tau\,\tau\dash},\tau / \tau\dash) 
\label{eqn:C11}
\\
 \removeLHspace
 \left<\! 
  \FT{S}\Sr_{\Phi}(\vect{k},\tau) \,
  \FT{S}_{\Psi}^{\mathrm{S}*}(\vect{k},\tau\dash) 
 \!\right>
 & \!\!\!=\!\!\! &
 \frac{\VEV^{4}}{\sqrt{\tau\,\tau\dash}} \frac{1}{V} 
 \FT{C}\Sr_{12}(k\sqrt{\tau\,\tau\dash},\tau / \tau\dash) 
\label{eqn:C12}
\\
 \removeLHspace
 \left<\! 
  \FT{S}\Sr_{\Psi}(\vect{k},\tau) \, 
  \FT{S}_{\Psi}^{\mathrm{S}*}(\vect{k},\tau\dash) 
 \right>
 & \!\!\!=\!\!\! &
 \frac{\VEV^{4}}{\sqrt{\tau\,\tau\dash}} \frac{1}{V} 
 \FT{C}\Sr _{22}(k\sqrt{\tau\,\tau\dash},\tau / \tau\dash).
\label{eqn:C22}
\end{eqnarray}
While the auto-correlations $\FT{C}\Sr _{11}$ and $\FT{C}\Sr _{22}$ are unchanged by the exchange of times $\tau\!\! \leftrightarrow \!\!\tau\dash$ (the scaling functions are real), the cross-correlation $\FT{C}\Sr_{12}$ is not. Applying this transformation to $\FT{C}\Sr _{12}$ gives the fourth member of this set, namely:
\begin{eqnarray}
 \removeLHspace
 \left<\! 
  \FT{S}\Sr_{\Psi}(\vect{k},\tau) \, 
  \FT{S}_{\Phi}^{\mathrm{S}*}(\vect{k},\tau\dash) 
  \!\right>
 & \!\!\!=\!\!\! &
 \frac{\VEV^{4}}{\sqrt{\tau\,\tau\dash}} \frac{1}{V}
 \FT{C}\Sr_{21}(k\sqrt{\tau\,\tau\dash},\tau / \tau\dash), 
\end{eqnarray}
and it is hence not independent: 
\begin{equation}
 \FT{C}\Sr_{21}(k\sqrt{\tau\,\tau\dash},\tau / \tau\dash)=\FT{C}^{\mathrm{S}^*}_{12}(k\sqrt{\tau\,\tau\dash},\tau\dash / \tau)
\end{equation}
This will be calculated for the present work since it will always be the case that $\tau \geq \tau\dash$ and hence the calculation of $\FT{C}\Sr_{21}$ is required for the cross-correlation to be fully determined.

In the vector case, the situation is slightly different in that the two independent degrees of freedom are spread across the three components of $\FT{S}\!\!\stackrel{^{V}}{}\!\!\!\!\!\!_{i}$. It would clearly be inefficient to consider the correlations between all such components and hence the procedure chosen here is to project the  two degrees of freedom by rewriting $\FT{S}\Vr{i}$ in terms of an orthonormal basis ($\hat{k}_{i},e^{1}_{i},{e}^{2}_{i}$):
\begin{equation}
 \FT{S}\Vr{i} 
 \; = \; \hat{k}_{i} (\hat{k}_{j}\FT{S}\Vr{j}) 
      + e^{1}_{i} \FT{S}^{\mathrm{V1}}
      + e^{2}_{i} \FT{S}^{\mathrm{V2}}
 \; = \; 
 e^{1}_{i} \FT{S}^{\mathrm{V1}}
 + e^{2}_{i} \FT{S}^{\mathrm{V2}}.
\end{equation}
The projection upon $\hat{\vect{k}}$ is a scalar and is therefore zero, which leaves the projections upon $\vect{e}^{1}$ and $\vect{e}^{2}$ as the desired vector degrees of freedom. It is hence not necessary to explicitly calculate $\FT{S}\!\!\stackrel{^{V}}{}\!\!\!\!\!\!_{i}$ at all, but merely to apply this procedure directly to $\FT{T}_{0i}$ and discard the scalar projection. 

Now, if all realizations are rotated $90^{\circ}$ about a particular $\vect{k}$, then the two projections change as: $\FT{S}^{\mathrm{V1}} \rightarrow \pm \FT{S}^{\mathrm{V2}}$, $\FT{S}^{\mathrm{V2}} \rightarrow \mp \FT{S}^{\mathrm{V1}}$. Hence statistical isotropy infers that their auto-correlations are equal:
\begin{eqnarray}
 \removeLHspace\!\!
 \left<\! 
  \FT{S}^{\mathrm{V1}}\!(\vect{k},\tau) \, 
  \FT{S}^{\mathrm{V1}^*}\!(\vect{k},\tau\dash) 
 \!\right>
 & \!\!\!=\!\!\! &
 \left<\! 
  \FT{S}^{\mathrm{V2}}\!(\vect{k},\tau) \, 
  \FT{S}^{\mathrm{V2}^*}\!(\vect{k},\tau\dash)  \!\right>
  \label{eqn:CV}
\\
 = & & \!\!\!\!\!\! 
 k^{2}\sqrt{\tau\,\tau\dash} \VEV^{4} 
 \frac{1}{V}
 \FT{C}\Vr{}\!\!(k\sqrt{\tau\,\tau\dash},\tau / \tau\dash),\nonumber 
\end{eqnarray}
but that their cross-correlation is zero. In the present case, involving a finite number of realizations, the two auto-correlations are averaged. The above scaling function definition matches that of Durrer et al. \cite{Durrer:1998rw, Durrer:2001cg}, although the basis projection approach here means that it is arrived at differently.

It should be noted that the definition of $\FT{C}\!\Vr{}$ differs from the general definition  by a factor of $k^{2} \tau \tau\dash$. This is relates to the fact that vector degrees of freedom can be written in terms of the curl of a vector field, and hence that $\FT{S}^{\mathrm{V}}_{i}$ measures angular momentum. Therefore, the UETC must decay for large scales. As noted in the introduction to this section, at low $k \sqrt{\tau \tau\dash}$ the scaling functions can be written as power laws expansions in $k \sqrt{\tau \tau\dash}$ and that they may contain only even terms. Hence in the vector case the scaling function may have a factor of $k^{2} \tau \tau\dash$ extracted so that all $\FT{C}$ tend to constants at low $k \sqrt{\tau \tau\dash}$ \cite{Durrer:1998rw, Durrer:2001cg}. 

Turning to the tensor modes, the source function $\FT{S}\Tr{ij}$ contains only two degrees of freedom and these may be projected out using a set of symmetric basis matrices:
\begin{equation}
 \FT{S}\Tr{ij} = M^{1}_{ij} \FT{S}^{T1} +  M^{2}_{ij} \FT{S}^{T2}.
\end{equation}
Considering the vector basis above, it may be noted that $k_{i} e^{A}_{i} e^{B}_{j}\!=\!0$ and $\delta_{ij} e^{A}_{i} e^{B}_{j}\!=\!\delta^{AB}$. Hence two symmetric matrices which project out tensor modes are:
\begin{eqnarray}
 M^{1}_{ij}
 & = &
 \frac{1}{\sqrt{2}} \left( e_{i}^{1} e_{j}^{2} + e_{i}^{2} e_{j}^{1} \right) 
\\
 M^{2}_{ij} 
 & = & 
 \frac{1}{\sqrt{2}} \left( e_{i}^{1} e_{j}^{1} - e_{i}^{2} e_{j}^{2} \right),
\end{eqnarray}
with the pre-factors required to make an orthonormal set. Consideration of a rotation by $45^{\circ}$ is sufficient to show that the cross-correlations are zero while the auto-correlations are again equal:
\begin{eqnarray}
 \removeLHspace\!\!
 \left<\! 
  \FT{S}^{\mathrm{T1}}\!(\vect{k},\tau) \, 
  \FT{S}^{\mathrm{T1}^*}\!(\vect{k},\tau\dash) 
 \!\right>
 & \!\!\!=\!\!\! &
 \left<\! 
  \FT{S}^{\mathrm{T2}}\!(\vect{k},\tau) \, 
  \FT{S}^{\mathrm{T2}^*}\!(\vect{k},\tau\dash)  \!\right>
\\
 = & & \!\!\!\!\!\! 
 2 \frac{\VEV^{4}}{\sqrt{\tau\,\tau\dash}} 
 \frac{1}{V}
 \FT{C}\Tr{}\!\!(k\sqrt{\tau\,\tau\dash},\tau / \tau\dash).\nonumber 
 \label{eqn:CT}
\end{eqnarray}
Note the factor of two, which is present to make $\FT{C}\Tr{}$ have the same definition as the tensor scaling function used by Durrer et al. \cite{Durrer:1998rw, Durrer:2001cg}.


\subsection{Eigenvector decomposition}
\label{subsec:eigendecomp}
Although the CMBEASY code produces the power spectra, the evolution of the perturbations is not performed using such quadratic quantities. Hence, although the UETC scaling functions defined above do contain all of the information required for CMB calculations, it is not immediately of the correct form for insertion into the source-enabled version of CMBEASY. 
However, suppose that a UETC set may be expanded as \cite{Turok:1996ud, Pen:1997ae}:
\begin{equation}
 \label{eqn:Udecomp}
 \FT{U}_{bc}(k,\tau,\tau\dash)
 = \sum_{n} \lambda_{n} 
    \FT{u}_{n}^{b}(k,\tau) 
    \FT{u}_{n}^{c}(k,\tau\dash).
\end{equation}
In the scalar case the indices $b$ and $c$ take on values 1 or 2, with  $\FT{U}\Sr_{11}$ formed from $\FT{C}\Sr_{11}$, but in the vector and tensor cases these indices are redundant. 
Now, the Green's function expression for the tensor component of a power spectrum (Eq.\ \ref{eqn:greens}) becomes:
\begin{equation}
 \label{eqn:greens2}
 \left<\! 
   \FT{X}_{a}(\vect{k},\tau) \FT{X}_{b}\conj (\vect{k},\tau)
 \!\right>
 =
 \sum_{n}
 \lambda_{n}
 I^{n}_{a}(k,\tau) \,
 I^{n}_{b}{\conj}(k,\tau), 
\end{equation}
where:
\begin{equation}
 I^{n}_{a}(k,\tau)
 =
 \int\! \diff\tau\dash \mathcal{G}_{a}(k,\tau,\tau\dash\!) \, \FT{u}_{n}(k,\tau\dash).
\end{equation}
As a result, the modified CMBEASY code can act upon the quasi-source functions $u_{n}^{c}$ and so calculate one term in the sum, and hence eventually, the desired power spectrum.

Since scaling is broken at the radiation-matter transition, then the UETC is given by a different scaling function within each era. However, suppose that the tensor scaling functions calculated under both radiation or matter domination can be similarly decomposed. Numerically, $\FT{C}(k\sqrt{\tau\,\tau\dash},\tau / \tau\dash)$ can only be represented at discrete values of its input parameters. Although it is not the form in which the simulations actually output data, a useful re-representation of the data is as $\FT{C}(k\tau , k\tau\dash)$, for a set of discrete $[k\tau]_{i}$ and $[k\tau\dash]_{j}$ with particular spacing. The scaling function is therefore represented as an $M\times M$ matrix $\FT{C}_{ij}$ and the decomposition is then: 
\begin{equation} 
 \label{eqn:Cdecomp}
 \FT{C}_{ij}
 = 
 \sum_{n} \lambda_{n} \; \FT{c}_{i}^{n} \; \FT{c}_{j}^{n}.
\end{equation}
Being real and symmetric $\FT{C}_{ij}$ is an Hermitian matrix whose eigenvectors form an orthonormal set. With that in mind, multiplication by $\FT{c}^{m}_{j}$ reveals that the supposed decomposition of $C_{ij}$ is nothing more than representing $\FT{C}_{ij}$ in terms of its eigenvectors $\FT{c}^{m}_{j}$ and eigenvalues $\lambda_{m}$:
\begin{equation}
 \FT{C}_{ij} \FT{c}_{j}^{m} = \lambda_{m} \FT{c}_{i}^{m}.
\end{equation}
In the continuum, eigenvectors become eigenfunctions, but the discussion is little different, while numerically the decomposition can be straightforwardly achieved using, for example, the in-built functions of MATLAB \cite{matlab}.

If the scaling functions in the radiation and matter eras are similar in form (but not necessarily in magnitude), then the resulting eigenvectors will be similar. Hence a valid (but approximate) means of dealing with the radiation-matter transition is to write the integral $I_{n}$ as \cite{Durrer:1998rw}:
\begin{equation}
 \sqrt{\lambda_{n}} I_{n} 
 \!=\! 
 \!\int\! \diff\tau\dash \mathcal{G}(k,\tau,\tau\dash\!) \, 
 \frac{\VEV^{2}}{\sqrt{\tau V}} 
 \!\!\sum_{\mathrm{x=r,m}}\!\!
  \alpha^{\mathrm{x}} \sqrt{\lambda_{n}^{\mathrm{x}}} \,
  \FT{c}^{\mathrm{x}}_{n}(k\tau\dash),
\end{equation}
where $^{\mathrm{r}}$ denotes the radiation era and $^{\mathrm{m}}$ the matter era. The function $\alpha^{\mathrm{r}}\!(\tau\dash)$ is equal to one deep in the radiation era and zero deep in the matter era, with a parameterizable transition between these limits as the dominant species changes (see \cite{Durrer:1998rw}). A similar approach is used to handle the transition from radiation to $\Lambda$ domination, although no scaling functions are calculated in this era and the defects sources are merely allowed to decay ($\alpha^{\mathrm{x}} \rightarrow 0$).

While this discussion is only marginally changed for the vector and tensor  contribution to the power spectrum, for the scalar modes it is complicated by the involvement of multiple UETCs and the non-symmetric cross-correlation $\FT{C}\Sr_{12}$. In this case a $2M\times2M$ matrix is formed as:
\begin{equation}
 C_{ij} = \left( \begin{array}{cc}
     \FT{C}_{11}\Sr & \FT{C}_{12}\Sr \\
     \FT{C}_{21}\Sr & \FT{C}_{22}\Sr
     \end{array} \right),
\end{equation} 
such that $C_{ij}\!=\!\FT{C}_{12}\Sr( [k\tau]_{i}, [k\tau]_{j-M} )$ if \mbox{$i\leq M$} but \mbox{$j>M$}. This matrix is then symmetric on account that $\FT{C}\Sr_{12}(k\tau,k\tau\dash)\!\!=\!\!\FT{C}\Sr_{21}(k\tau\dash,k\tau)$ and these appear off-diagonal. From Eqs. \ref{eqn:Udecomp} and \ref{eqn:Cdecomp}, the first half of the resulting eigenvectors then represent the $\FT{S}_{\Phi}$, while the second half replaces $\FT{S}_{\Psi}$. 

Although CMBEASY does not actually involve such a Green's function approach, this eigenvector decomposition allows the UETC data to be incorporated and the corresponding CMB power spectrum calculated as the sum of eigen-contributions.


\subsection{Collection of UETC data}
As already noted, the numerical simulations start from random initial conditions and therefore do not start in the scaling regime, or in fact even contain strings at very early times. Hence there is an initial phase during which UETC data cannot be taken. Assuming that the system  tends towards a scaling solution, eventually a time $\tau_{\mathrm{scaling}}$ is passed at which the system scales to within the desired accuracy. The approach chosen here is to then store the six 3D arrays corresponding to the variables $\FT{S}^{\mathrm{S}}_{\Phi}$, $\FT{S}^{\mathrm{S}}_{\Psi}$, $\FT{S}^{\mathrm{V1}}$, $\FT{S}^{\mathrm{V2}}$, $\FT{S}^{\mathrm{T1}}$ and $\FT{S}^{\mathrm{T2}}$ at that time. Then for each output time \mbox{$\tau \geq \tau_{\mathrm{scaling}}$}, correlations are performed against these reference data. Angular averaging is performed (as well as the averaging of the two vector and two tensor UETCs) and the scaling functions $\FT{C}\Sr_{11}$, $\FT{C}\Sr_{12}$, $\FT{C}\Sr_{21}$, $\FT{C}\Sr_{22}$, $\FT{C}\Vr{}$ and $\FT{C}\Tr{}$ output to disk as six 1D arrays corresponding to the time ratio $\tau / \tau_{\mathrm{scaling}}$. As previously noted, the output of $\FT{C}\Sr_{21}$ is required since data for ratios less than unity are not calculated.

Since the simulation is expected to be closest to scaling at later times, there is an argument for performing all unequal-time correlations against the final simulation time: $\FT{U}(k,\tau,\tau_{\mathrm{end}})$. However, this means that the information required for correlation is not available until the simulation has been evolved to $\tau_{end}$ and the data from every prior output time must either be stored, or part of the simulation re-ran once the reference data has been acquired at $\tau_{\mathrm{end}}$. Given that the 3D arrays required for a $512^{3}$ lattice need \mbox{3 GB} of storage, then storing them for each output time would be inefficient, while the second option requires the repetition of a significant proportion of the calculation. Additionally, it is the desire to run the simulations beyond their strict causal limit, in which case it is preferable to use such times merely for the less important UETC data from the most extreme time ratios, rather than for the important equal-time data. 


\section{Simulation and UETC results}
\label{sec:results}

\subsection{Simulations involved}
The results that are described here stem from \mbox{$15\,000$ CPU-hours} of calculation on the UK National Cosmology Supercomputer 
\cite{cosmos-website}, using $512^{3}$ lattices and run across $64$ CPUs. This enabled 5 realizations in both radiation and matter eras for $s\!=\!0.0$, $0.2$ and $0.3$ and, additionally, $s\!=\!1$ in the radiation era. At larger $s$ values, the network formation is more challenging and the initial period of non-scaling is generally larger. The largest value of $s$ for which scaling could be attained quickly enough for the required UETC data to be taken from a matter era simulation was $s\!=\!0.3$ and this is hence the closest value to the $s\!=\!1$ goal. The $s\!=\!1$ runs under radiation domination provide useful data on the non-artificial case, but scaling was not reached fast enough for all of the required UETC data to be collected and these are incorporated merely for reference. 

While of course this situation could be improved using larger \mbox{$(512\times M)^{3}$}  lattices, which allow for longer causal run times, the processing time is very sensitive to such enlargements and varies as $M^{4}$. On the other hand, the actual increase in the available $s$ values in the matter era would have been slight and in fact, from the results obtained, there is little to justify such a large computational outlay. 


\subsection{Scaling in the string length}
It is obviously of great importance to test that the system approaches a scaling regime, and to a satisfactory level prior to the acquisition of UETC data. Further, the most important such test necessarily involves the actual UETC itself, via which scaling is actually employed here. However as an auxiliary test, we additionally use the behaviour of the mean string length $L_{\mathrm{H}}$ in each horizon volume $V_{\mathrm{H}}$. Given scaling, $L_{\mathrm{H}}$ grows in proportion to the horizon size $\tau$ while $V_{\mathrm{H}}$ varies as $\tau^{3}$, resulting in:
\begin{equation}
 \xi = \sqrt{ \frac{V_{\mathrm{H}}}{L_{\mathrm{H}}} } \propto \tau.
\end{equation}

In the context of idealized-string simulations, this is expressed in terms of the Lorentz-invariant length and a simulation segment usually represents an element of this. In the present context the length must be derived from the fields and there is no unique means of proceeding. Vincent et al. \cite{Vincent:1997cx} and Moore et al. \cite{Moore:2001px} use a net winding of the phase around the smallest closed loops represented on the lattice, that is around lattice-plaquettes, in order to detect strings penetrating them. This enables the string paths to be traced and so the string length estimated. Here we employ a simpler approach and use the mean Lagrangian density in order to estimate the invariant length:
\begin{equation}
 \mu \frac{L_{\mathrm{H}}}{V_{\mathrm{H}}}
 \; = \;
 - \mathcal{L}
\end{equation}
(since $-\mathcal{L}$ is the energy density for a static string, vanishes for perturbative radiation, and is a 4-scalar). Although this is in many ways inferior to the above method, the results will be needed in the next sub-section and it will then be desirable that they directly involve the entire simulation volume.

\begin{figure}
\resizebox{\columnwidth}{!}{\includegraphics{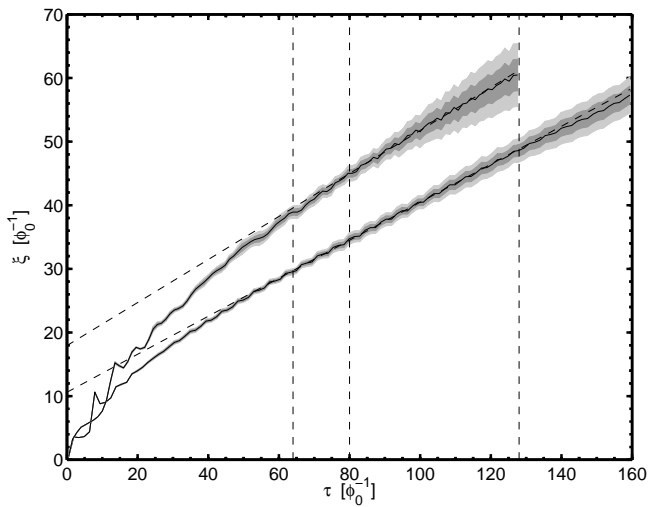}}
\caption{\label{fig:xi}Results for the Lagrangian measure of $\xi$ from 5 simulations in the radiation era with $s\!=\!1$ (top) and $s\!=\!0$ (lower). The shaded regions show the $1\!-\!\sigma$ and $2\!-\!\sigma$ variations in the mean (error bars have not been used since correlations extend across most of the plot and would tempt the reader into believing individual points were independent). The best-fit straight lines over the region $80\VEV^{-1}<\tau<128\VEV^{-1}$ \mbox{($s\!=\!1$)} and $64\VEV^{-1}<\tau <128\VEV^{-1}$ \mbox{($s\!=\!0$)} are also shown. Note that for the later case, this excludes the acausal final period.}
\end{figure}
The results from 5 simulations in the radiation era with $s\!=\!1$ and $s\!=\!0$ are presented in Fig.\ \ref{fig:xi}. The best-fit lines show a linear regime is reached of form:
\begin{equation}
 \xi
 \; \propto \;
 (\tau - \tau_{\xi=0}).
 \label{eqn:XiOffset}
\end{equation}
Similar results are found for the other simulations, including those in the matter era, with best-fit gradients as shown in Table \ref{tab:gradients}. It is encouraging that the slopes show no resolvable trend with $s$ in either era and that the $s\!=\!0$ case appears to be a good approximation to $s\!=\!1$ under radiation domination (as was found previously by Moore et al. \cite{Moore:2001px}). Further, in the $s<1$ cases, which were ran with a final acausal period to extend the $\tau /\tau\dash$ range, the trend continues to within the statistical uncertainties.
\begin{table}
\begin{ruledtabular}
\begin{tabular}{ccc}
 Simulation set & \multicolumn{2}{c}{$\xi$ versus $\tau$ gradient} \\
         & radiation       & matter \\
 \hline
 $s=1.0$ & $0.33  \pm 0.02  $         \\
 $s=0.3$ & $0.299 \pm 0.014 $ & $0.33 \pm 0.03$   \\
 $s=0.2$ & $0.299 \pm 0.012 $ & $0.304 \pm 0.012$ \\
 $s=0.0$ & $0.31  \pm 0.02  $ & $0.304 \pm 0.013$  \\
\end{tabular}
\end{ruledtabular}
\caption{\label{tab:gradients}The gradient of the linear region of the $\xi$ versus $\tau$ plots from 5 realizations in each case. Note that results in this table do not involve the acausal over-run period.}
\end{table}

All runs show this offset scaling behaviour including smaller runs which employed a gauge-invariant \cite{Kajantie:1998bg} (but unsmoothed) version of the above phase-winding approach to the string length. It is hence not a result of the string length measure employed but merely a consequence of the initial period of non-scaling. Choosing a different network formation history (that is a different $\tau$ dependence of $s$ at early times) gives no detectable change in the slopes, but has an effect on the $\tau_{\xi=0}$ values. Hence, there is nothing fundamental about the offset and it is only the gradients that are important. Note that if the apparent linear regime was false, then it would be expected that the gradient too would be dependent upon the formation history (and therefore $s$ as the early $s(\tau)$ dependence must then be different). However, this is not the case. 

Further, if the seen trends are extrapolated to times of cosmological relevance, then the fractional disagreement between this and a direct proportionality becomes insignificant and therefore these results are believed to be evidence for the required scaling behaviour. 


\subsection{Scaling in the equal-time scaling functions}
\begin{figure}
\resizebox{\columnwidth}{!}{\includegraphics{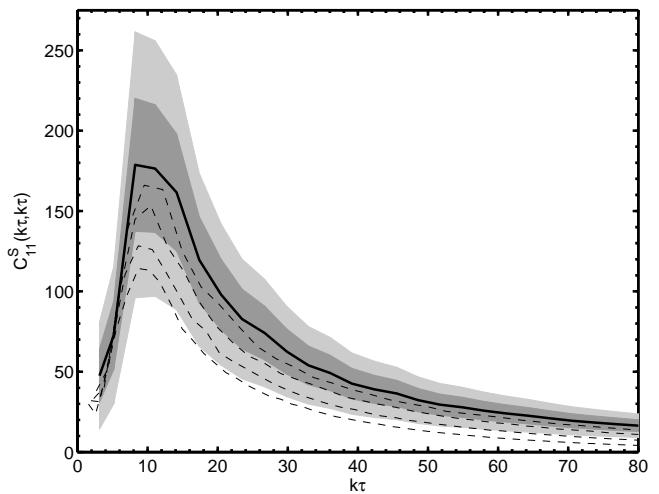}}
\caption{\label{fig:ETCbad}The raw equal-time scaling function $\FT{C}\Sr_{11}(k\tau,k\tau)$ as averaged over 5 realizations for $s\!=\!0.3$ in the radiation era. Results are plotted at roughly uniformly-spaced $\tau$ values in the range \mbox{$64\VEV^{-1}<\tau<128\VEV^{-1}$}. The lower lines correspond to increasingly early times (dashed), with the $1\!-\!\sigma$ and $2\!-\!\sigma$ uncertainties in the mean indicated for the latest time (solid).}
\end{figure}
\begin{figure}
\resizebox{\columnwidth}{!}{\includegraphics{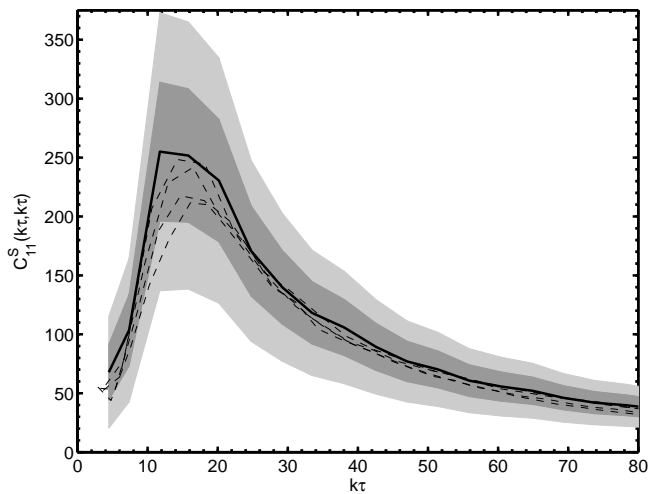}}
\caption{\label{fig:ETCgood}The equal-time scaling function $\FT{C}\Sr_{11}(k\tau,k\tau)$
as in the previous figure, but with the time offset taken into account. For present plot, the
mean offset across the 5 realizations is used to adjust results for each one, whereas the actual CMB calculations use independent offsets for each realization.} 
\end{figure}
A more complete test of the scaling hypothesis is afforded by the UETC scaling functions, which should show no absolute temporal dependence, being a function merely of $k\sqrt{\tau\,\tau\dash}$ and $\tau/\tau\dash$. This provides an analysis as a function of scale and this is the test of direct relevance for the present CMB calculations. Since the functions are most important for equal (or near-equal) times, then a useful test is to compare the equal-time scaling functions calculated at different $\tau$. Typical such results are shown in Fig.\ \ref{fig:ETCbad} for $\FT{C}\Sr_{11}$, during the period \mbox{$64\VEV^{-1}<\tau<128\VEV^{-1}$} when $\xi$ varies linearly with $\tau$ in the $s<1$ cases. 

Unfortunately, a systematic variation is clearly visible such that the overall normalization increases with $\tau$, while the peak becomes broader and shifts to the right. However, if all times in the calculation of $\FT{C}$ are replaced as:
\begin{equation}
 \tau = (\tau_{\mathrm{sim}} - \tau_{\xi=0}),
\end{equation}
then the effect is a symmetric rescaling of the axes by a factor dependent upon the simulation time $\tau_{\mathrm{sim}}$. This may be thought of as adjusting the results to estimate the scaling functions at late times, when the correction would have no effect. Once this process is performed, there is the desired time-constancy of the scaling functions, as is shown in Fig.\ \ref{fig:ETCgood} \footnote{The scaling function plots presented here involved a common offset within each set of runs while the actual CMB calculations will incorporate independent offsets for each realization. The latter means that each realization provides outputs at different effective times such that the mean and spread calculations required for these plots would have necessitated interpolations, while all realizations would not have provided data for the earliest and latest times. Given the already short range of times over which scaling may be examined and a desire to keep the results presented as raw as possible, independent offsets were not used until the eigenvector decomposition phase, when the interpolation was already needed in order to transfer the UETC data onto the matrix grid.}. Similar results are seen for the other scaling functions and $s$ values, but with slight exception of $s\!=\!1$. In this more ambitious case, good scaling is not seen for simulation times as early as $\tau_{\mathrm{sim}} \approx 64 \VEV^{-1}$ and reliable UETC data cannot be taken until later in the simulation. 

\begin{figure}
\resizebox{\columnwidth}{!}{\includegraphics{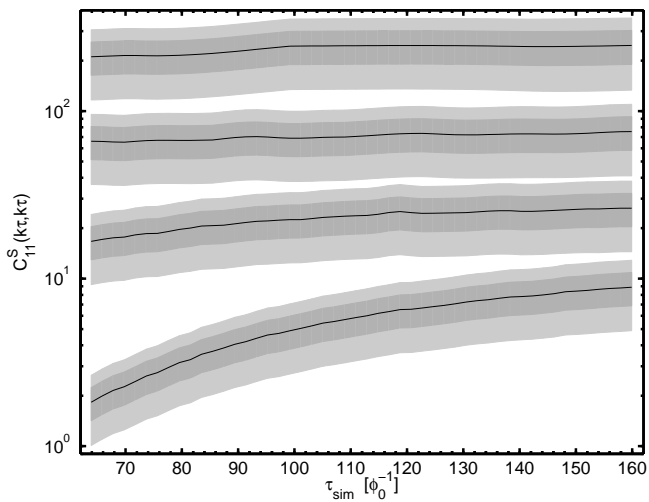}}
\caption{\label{fig:ETCconverge}The temporal dependence of the offset corrected $\FT{C}\Sr_{11}(k\tau,k\tau)$ at four fixed values of $k\tau$: $17$, $50$, $110$ and $220$ (which appear top $\rightarrow$ bottom in the plot). Results are from 5 realizations with $s\!=\!0.3$ in the radiation era, using a common time offset for each realization.}
\end{figure}
This behaviour may be further explored by plotting the $\FT{C}(k\tau,k\tau)$ for a given $k\tau$ as a function of time, as in Fig.\ \ref{fig:ETCconverge}. The results are qualitatively similar for the other scaling functions and $s$ values, such that the important scales, close to the horizon, show scaling to a good approximation before UETC data acquisition begins for the $s<1$ cases ($\tau_{\mathrm{sim}}\!=\!64\VEV^{-1}$). However, scales which are small relative to the horizon take longer to show convergence. Clearly this must be the case at some level in the present simulations since the string width is around one sixtieth of the horizon size, or larger, when UETC data is first collected. However, the use of such a dataset for CMB calculations should be accurate, considering the statistical uncertainties, given that the non-scaling effects at high $k\tau$ are unlikely to be larger than the statistical uncertainties at more important values. Hence, it would seem that the required UETC data is provided by the present $s\!<\!1$ simulations with $512^{3}$ lattices. 

\subsection{Dependence of the ETCs upon $s$}

\begin{figure*}
\resizebox{\columnwidth}{!}{\includegraphics{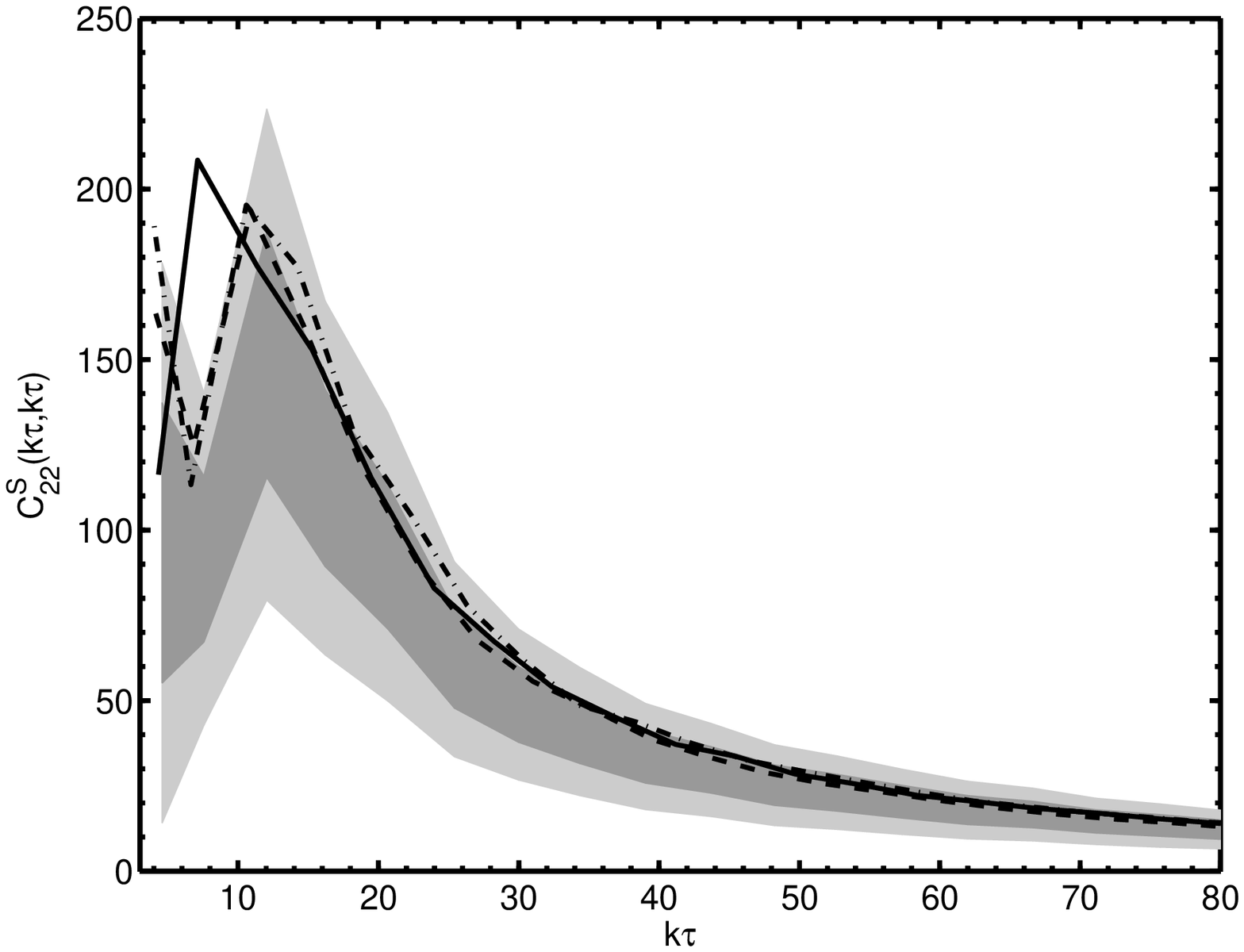}}
\resizebox{\columnwidth}{!}{\includegraphics{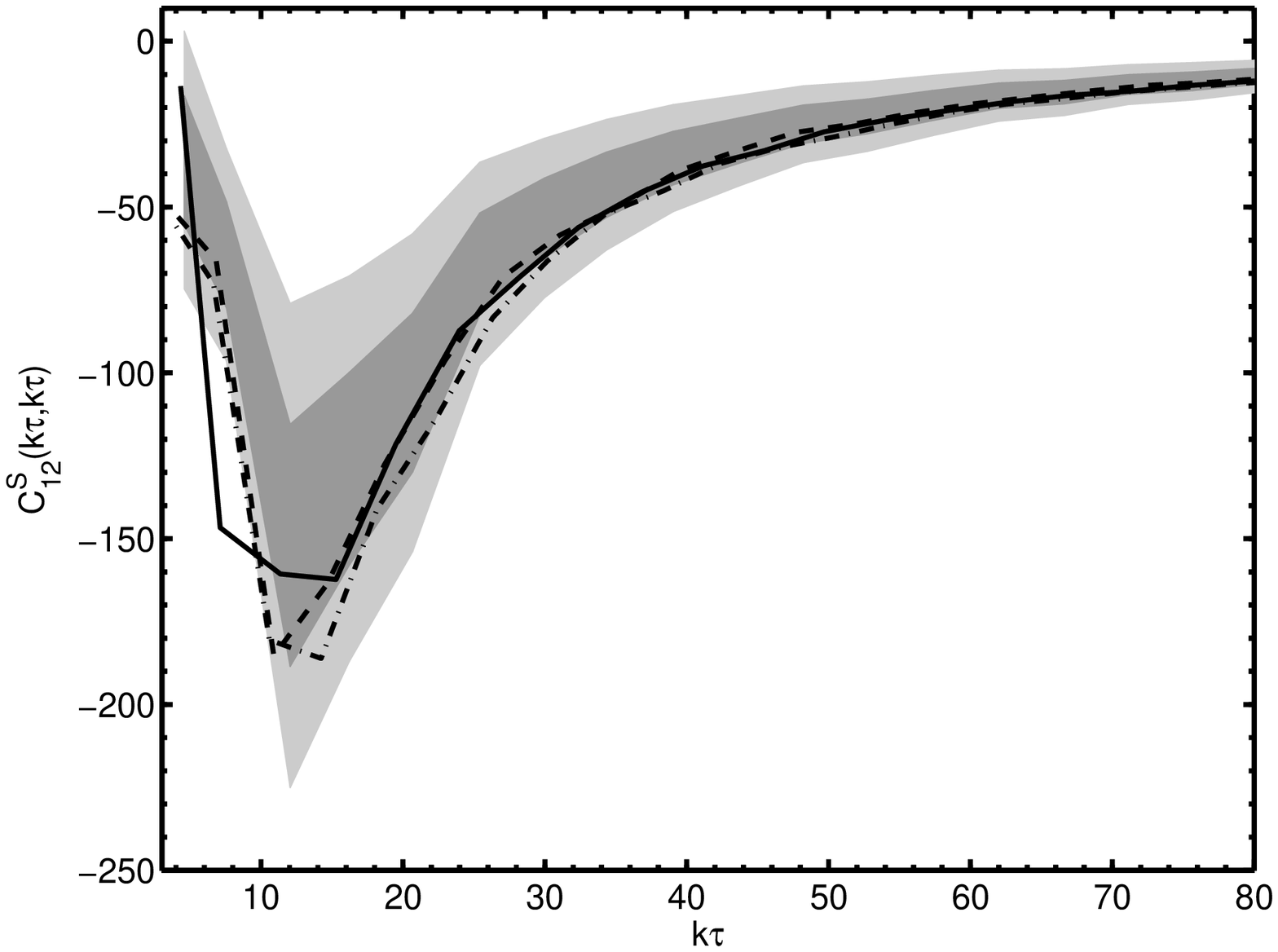}}
\resizebox{\columnwidth}{!}{\includegraphics{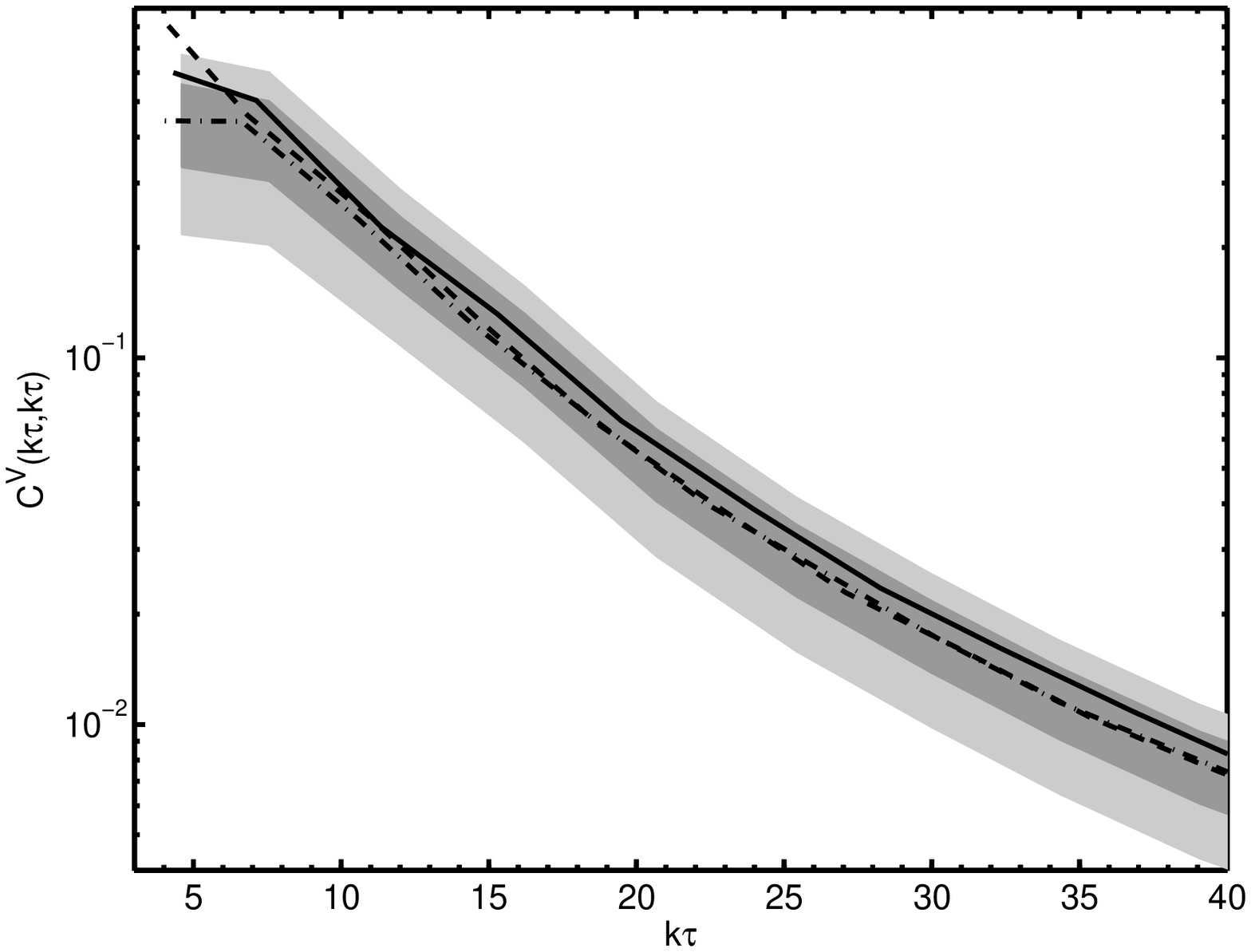}}
\resizebox{\columnwidth}{!}{\includegraphics{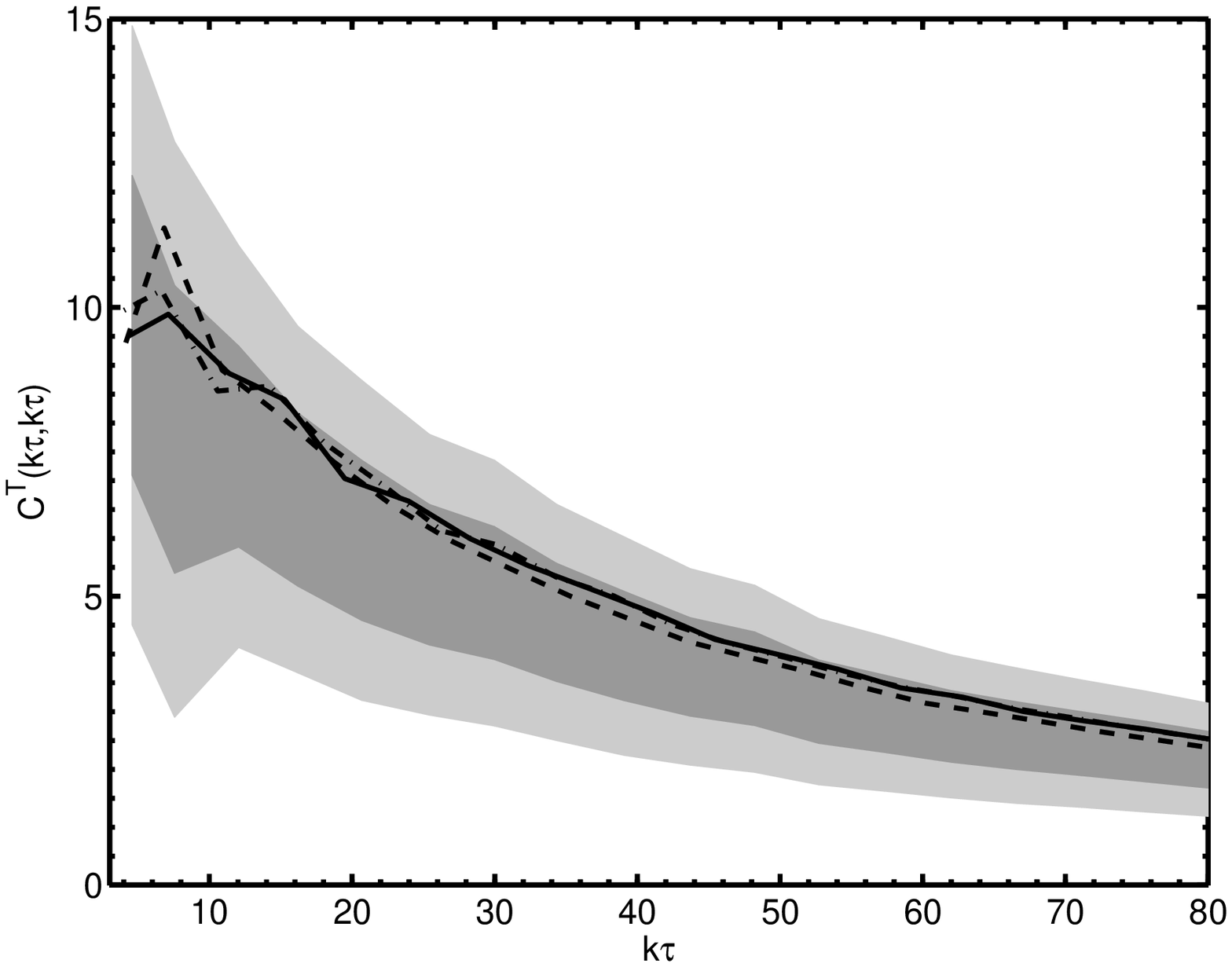}}
\caption{\label{fig:ETCvaryS}The dependence of the $\FT{C}(k\tau,k\tau)$ upon $s$ in the radiation era at $\tau_{\mathrm{sim}}\!=\!128\VEV^{-1}$. The shaded regions show the $1\!-\!\sigma$ and $2\!-\!\sigma$ uncertainties for the $s\!=\!1$ case, while the lines indicate the $s\!=\!0.3$ (solid), $s\!=\!0.2$ (dashed) $s\!=\!0.1$ (dot-dash) results.}
\end{figure*}
The dependence of such equal-time results upon $s$ are shown in Fig.\ \ref{fig:ETCvaryS} for radiation era. This figure shows results from the end of the causal period of simulation, by which time even the true $s\!=\!1$ case shows good scaling. The results highlight that there is at most a very weak dependence upon $s$ and that the typical difference between $s\!=\!1$ and the artificial cases is approximately equal to the statistical uncertainties. Although an $s\!=\!1$ reference is not available in the matter era there are no obvious trends between the $s\!<\!1$ cases for the scaling functions in that era, just as in the included figure. 


\subsection{Unequal-time scaling function results}

\begin{figure*}
\resizebox{\textwidth}{!}{\includegraphics{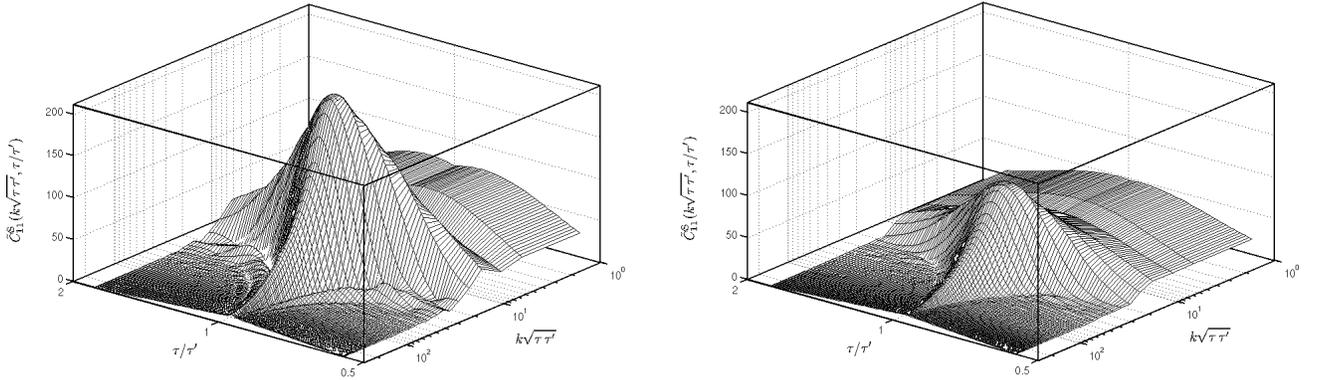}}
\caption{\label{fig:UETCscalar}The unequal-time scaling functions $\FT{C}\Sr_{11}$ in the
radiation era (left) and matter era (right). Results are from 5 realizations with $s\!=\!0.3$. Each realization is given the average offset time such there is no interpolation involved which might otherwise smooth these plots. The data is then raw, albeit it has been extrapolated as a constant for low $k\sqrt{\tau\,\tau\dash}$ and the relative time symmetries have been
used since the simulations output only for $\tau > \tau\dash$. Note there are large statistical uncertainties which cannot be shown here.}
\end{figure*}

\begin{figure*}
\resizebox{\textwidth}{!}{\includegraphics{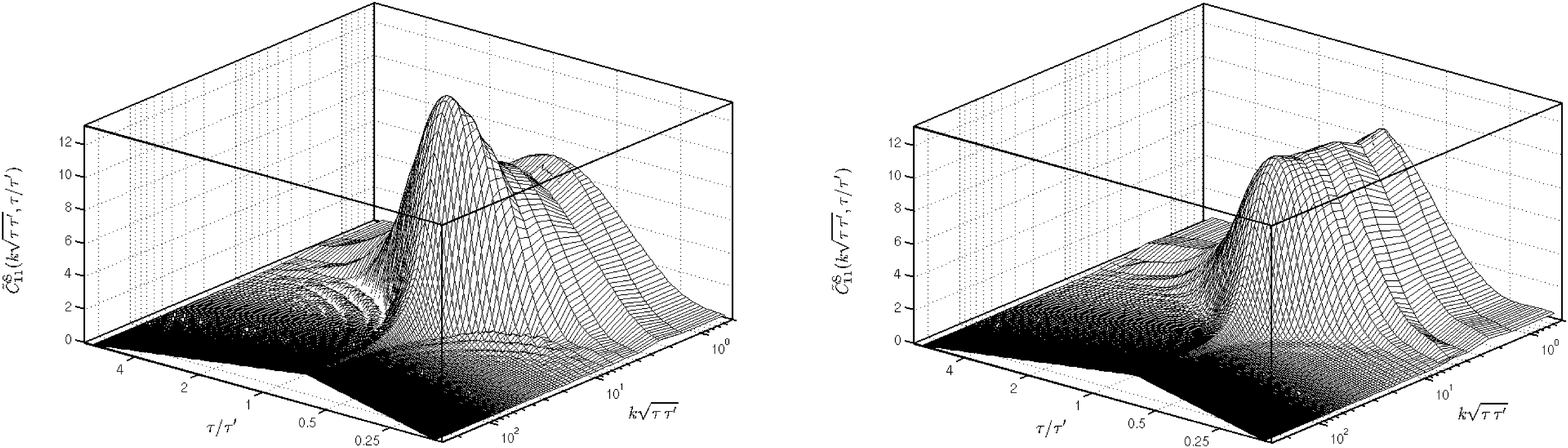}}
\caption{\label{fig:textureUETC}The unequal time scaling functions from a global texture model, in order to provide a comparison case for the previous figure. The results show are from 10 realizations in the radiation era (left) and matter era (right), using a $256^{3}$ lattice. Note the difference in the axis scales between this and the string case.}
\end{figure*}

Although the equal-time data does exhibit scaling, it is additionally required that the simulations are able to sample all of the important regions of the unequal-time scaling functions and, for example, the available $\tau/\tau\dash$ is sufficient. This ratio is clearly constrained by the late onset of scaling, which is significantly later than the case of a non-linear $\sigma$-model, which has been previously used to represent global defects in field simulations \cite{Pen:1993nx, Pen:1997ae, Durrer:1998rw}. Here, we have additionally simulated global O(4) textures via a very similar procedure to those previous works, in order to provide a useful reference with which to make comparisons against, but also as a check of the UETC and CMB calculation algorithms. In the texture case, the scaling regime may be comfortably studied in smaller $256^{3}$ simulations, with a maximum reliable time ratio of $\approx\!6$. In comparison, the larger string simulations can provide a maximal ratio of $160/64\!=\!2.5$ \mbox{($s\!<\!1$)}, which is then further reduced when the time offset is applied, becoming slightly less than $2$ and dependent upon the precise case in question.

The form of $\FT{C}\Sr_{11}$ under both radiation and matter domination is shown for strings in Fig.\ \ref{fig:UETCscalar}. Results are shown for the $s\!=\!0.3$ case but there are very similar for the other $s$ values simulated: there is pronounced peak for $\tau\!=\!\tau\dash$ and $k\tau\!\approx\!17$, with a decline for unequal-times or for large $k\sqrt{\tau\tau\dash}$ values. Unfortunately, the late onset of scaling means that the super-horizon plateau is not well-covered by these simulations. It is hence shown here via an extrapolation at a constant level, determined from the mean of the function at each $\tau/\tau\dash$ for the lowest $k\sqrt{\tau\tau\dash}$ values present. Further, it should be stressed that large statistical uncertainties exist which cannot be well-shown on these 3D plots. 

\begin{figure*}
\resizebox{\textwidth}{!}{\includegraphics{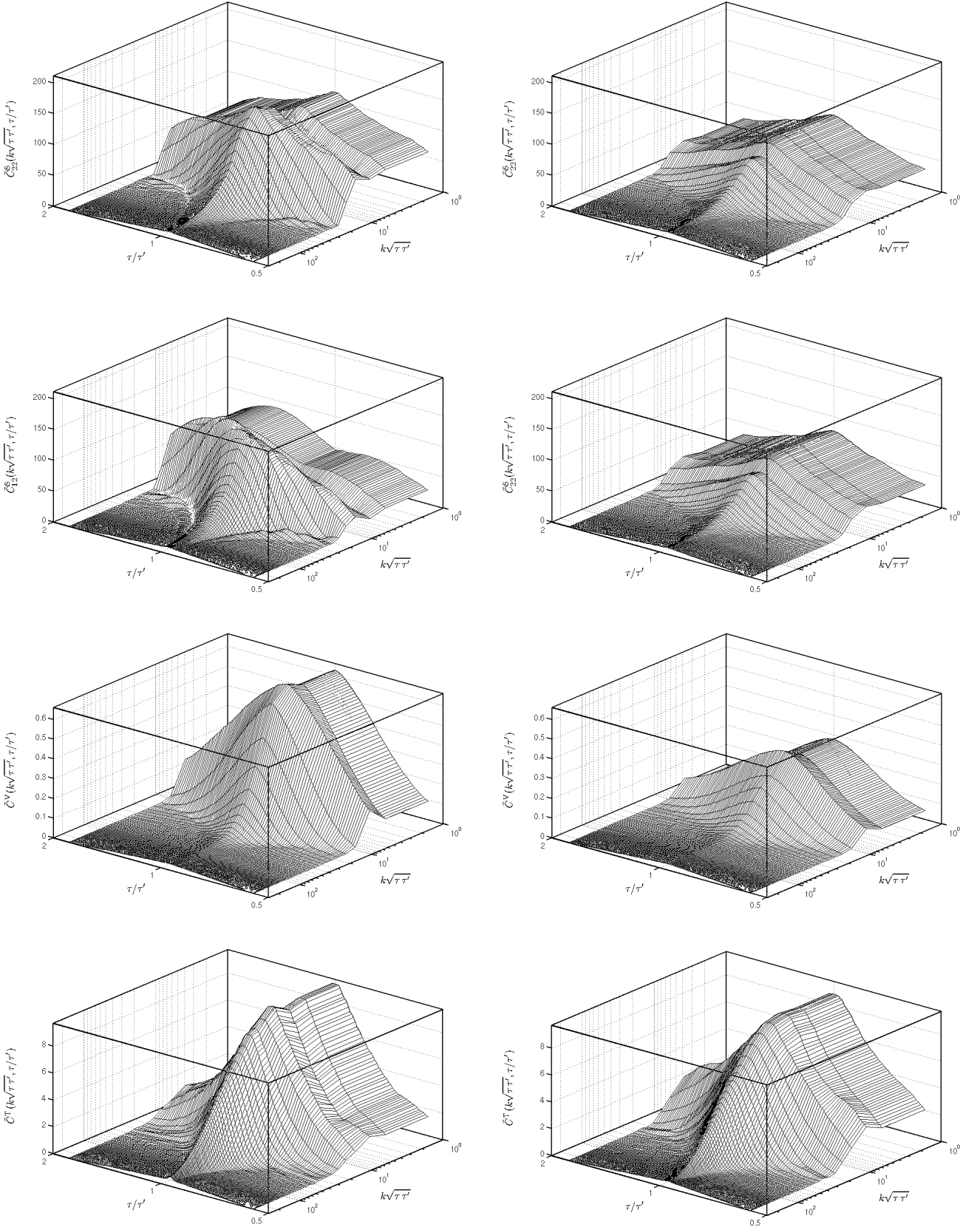}}
\caption{\label{fig:UETCs}The remaining unequal-time correlation functions from $s\!=\!0.3$ string simulations. For the cross correlation function $\FT{C}\Sr_{12}$ only the magnitude is shown and by virtue of Eq.\ \ref{eqn:defPsi} this is almost entirely anti-correlated. Note that its form can vary significantly from realization-to-realization and hence it suffers from
very large statistical uncertainties. The vertical axes are constant within each scalar-vector-tensor class (including Fig.\ \ref{fig:UETCscalar}) so as to aid size comparisons.}
\end{figure*}

For comparison, the corresponding results from the texture model are shown in Fig.\ \ref{fig:textureUETC}. It is immediately apparent that while the string case is dominated by a peak in both eras, this is not so in the texture case, which involves a merely small peak above the plateau in the radiation era (with a possible smaller and poorly resolved version under matter domination). However careful attention should be paid to the axes, which reveal that the string scaling functions are significantly larger than those of textures. Further, the string case shows significant contributions at relatively large $k\sqrt{\tau\,\tau\dash}$ values, whereas in the texture case $\FT{C}\Sr_{11}$ decays for $k\tau \gtrsim 10$ ($\tau\!=\!\tau\dash$). Cosmic strings are difficult structures to remove and while the highly curved super-horizon forms quickly straighten within the horizon, so that their length is very-much reduced, strings do persist inside the horizon for some time. On the other hand, texture configurations may be continuously smoothed away such that they are less significant on smaller scales. This difference is highlighted by the sub-horizon peak in this string scaling function, which corresponds roughly to the inter-string separation. Clearly, these differences will manifest themselves in the resulting CMB predictions. 

The other unequal-time scaling functions from the $s\!=\!0.3$ string simulations are shown in Fig.\ \ref{fig:UETCs}. Each share an increase in magnitude and a shift to small scales relative to the texture case, although the corresponding plots are not shown due to brevity considerations. With the exception of the tensor functions, they are larger under radiation domination than in the matter era. However, the overall forms do not change greatly between the two eras, which is desirable given the interpolation of the sources used in order to model the radiation-matter transition. The differences are hence largely associated with the eigenvalues rather than changes in the forms of the eigenvectors.

With regard to the question of $\tau/\tau\dash$ coverage, all but $\FT{C}\Sr_{22}$ and $\FT{C}\Sr_{12}$ are clearly well-sampled, with significant decay at unequal times. It should be noted, however, that the cross-correlation function suffers from very large realization-to-realization differences and that the systematic uncertainties arising from incomplete coverage of this function are likely to be insignificant relative to those of a statistical nature. This is also likely to be the case for the other functions and such systematic effects will be estimated as part of the CMB calculation. 


\section{CMB power spectra}
\label{sec:CMBPowSpe}

\begin{figure}
\resizebox{\columnwidth}{!}{\includegraphics{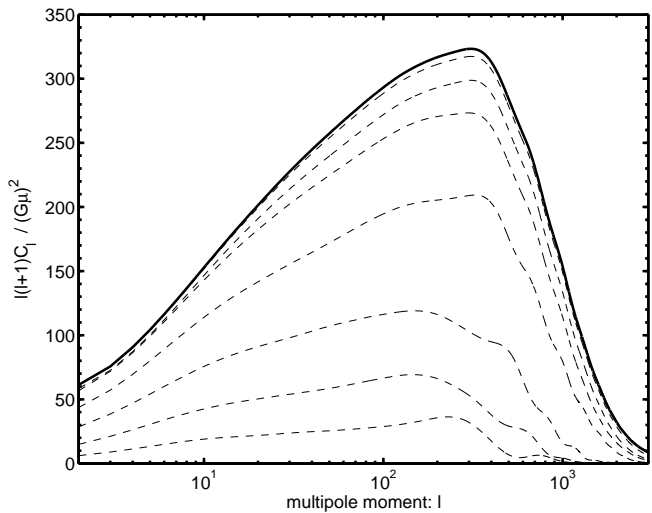}}
\caption{\label{fig:TTconverge}The convergence of the eigen-contribution sum for the temperature power spectrum due to cosmic strings. Results are for $s\!=\!0.3$ with (dashed) 1, 2, 4, 8, 16, 32, 64 and (solid) 128 terms included.}
\end{figure}

After the eigenvector decomposition of the above results was performed, using a matrix size $M\!=\!512$ and independent offsets for each realization, the modified version of CMBEASY was applied. This is the only point at which the cosmological parameters are involved and these were chosen to match the central values from non-CMB determinations: $h\!=\!0.72\pm0.08$ \cite{Freedman:2000cf}, $\Obhh\!=\!0.214\pm0.0020$ \cite{Kirkman:2003uv}, $\Ol\!=\! 0.75^{+0.06}_{-0.07}$ \cite{Knop:2003iy}; with additionally the inflation-motivated assumption of spatial flatness (and $0.1$ used as the optical depth to the last-scattering surface). The string contribution to the temperature power spectrum is then given by the sum over eigen-contributions, each of which took of order $40$ minutes of calculation time on a 2.4GHz 64-bit AMD Opteron. This process is somewhat slower than the CMBEASY calculation of an inflation power spectrum since the vector mode must be additionally evolved, but also because the oscillating source functions necessitate a more careful integration of the equations. The convergence of the eigen-sum in the $s\!=\!0.3$ case is shown in Fig.\ \ref{fig:TTconverge}, with truncation after $128$ terms giving convergence to within one percent.

\begin{figure}
\resizebox{\columnwidth}{!}{\includegraphics{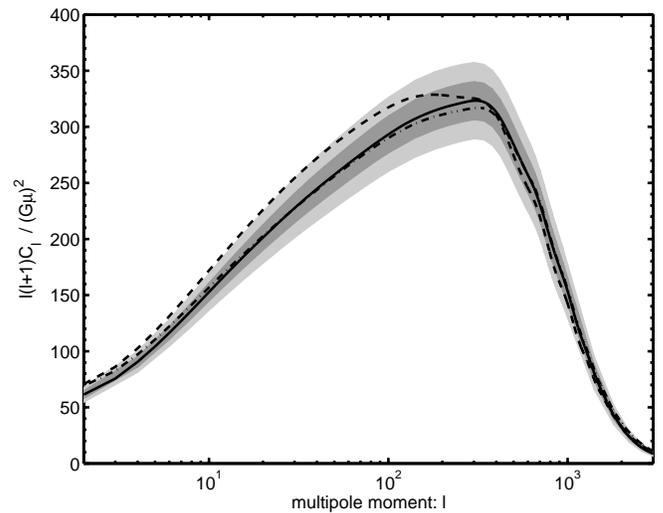}}
\caption{\label{fig:TTvaryS}The CMB power spectrum contribution from cosmic strings simulated
with $s=0.3$ (solid), $0.2$ (dashed) and $0.0$ (dot-dashed). The estimated $1\!-\!\sigma$ and $2\!-\!\sigma$ uncertainties (in the mean) are indicated in the $s=0.3$ case by the shaded regions.}
\end{figure}

We estimate the statistical uncertainties in these power spectra by repeating the CMB calculations using UETC results from individual realizations rather than first averaging the scaling functions. These are then shown in Fig.\ \ref{fig:TTvaryS} to be somewhat larger than the truncation error. Note however, that the statistical uncertainties are not directly related to cosmic variance since these results use the scaling approximation and do not stem from statistical variations across a volume that actually corresponds to the observable universe. 

Additional systematic uncertainties stem from the limited $\tau/\tau\dash$ ratios achievable in the string simulations during the scaling regime. No extrapolation was made for the results presented here and the correlators were taken to be zero where data was unavailable. The corresponding systematic uncertainty can be explored by further zeroing all correlator data beyond a certain $\tau/\tau\dash$ value, with the trend suggesting that the power spectra results are perhaps of order $10\%$ too high as a result of the limited $\tau$-range available, a result which is insensitive to the particular $\ell$ value. Finite-volume effects relating to the fact the system is only scaling approximately (particularly at earlier times in one or two realizations) can be explored by performing the $\xi$ fit over only a sub-set of the $\tau_{\textrm{sim}}=64\rightarrow128$ interval. It is possible that the power spectra are underestimated by of order $10\%$, which again is not heavily dependent upon $\ell$. Other systematic sources of error include the matrix re-representation of the UETC data and associated numerical errors but a halving of the matrix size, for example, has a negligible effect on the results. The use of logarithmic rather than linear spacing in the $k\tau-k\tau^{\prime}$ plane slows the convergence of the eigen-contribution sum significantly, but upon convergence gives merely a slight shift on the high $\ell$ of the peak and a change in the numerical results of $10\%$ for $\ell=1000$ but less than one percent for $\ell=10$, $100$ or $300$. The modelling of the matter-radiation transition is a further source of uncertainty, for which an overestimate can be derived by the use of the matter scaling functions in the radiation era also. The contribution to large scales comes almost exclusively from the string sources in the matter era and the corresponding change in the power is neglible for $l\sim10$. The lower normalization of the matter era UETCs gives a reduction in power of $4\%$ for $\ell=100$, which increases to $14\%$ for $\ell=1000$, but these are clearly greater than the actual modelling errors. 

Fig.\ \ref{fig:TTvaryS} additionally explores the effect of a change in $s$, which is of particular importance. It was initially anticipated that variations in $s$ would provide a trend via which an extrapolation to the $s\!=\!1$ result could be made. However, the $\xi$ and equal-time correlation results showed that any such variation was likely to be at most comparable to the statistical uncertainties, even for full range $s\!=0\!\rightarrow\!1$ that was explored under radiation domination. The power spectrum results are consistent with this conclusion and it appears that there is no basis upon which to perform any such extrapolation, with the variation from $s\!=\!0.0$ to $s\!=\!0.2$ being reversed for the jump to $s\!=\!0.3$ and falling within the estimated statistical uncertainties. The position must therefore be taken that the present use of $s\!<\!1$ simulations instills merely a systematic uncertainty in the results, which is of a magnitude comparable to or smaller than the statistical uncertainties. For example, under a linear fit for the variation with $s$, the data of course allows for a zero gradient, but suggests an $s=1$ extrapolation value that is $5\%$ greater than that at $s=0.3$ for $\ell=10$, although only $0.4\%$ is due to the gradient, with the offset dominating. We refer the reader to the appendix for a more detailed discussion of the systematic
uncertainties.

\begin{figure}
\resizebox{\columnwidth}{!}{\includegraphics{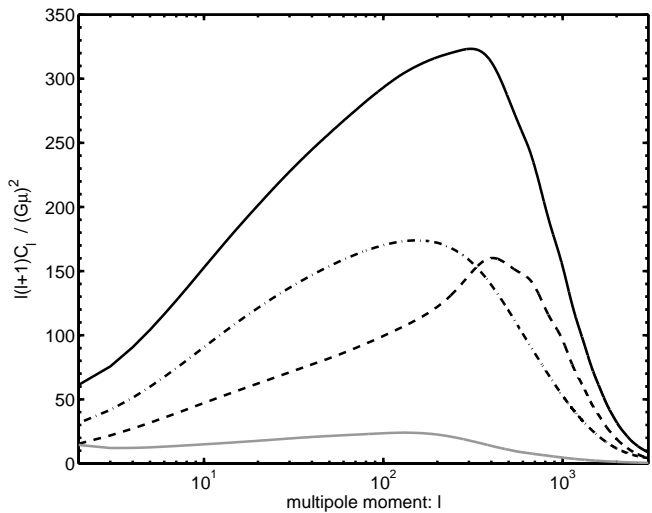}}
\caption{\label{fig:TTdecomp}The decomposition of the CMB power spectrum (solid) into scalar (dashed), vector (dot-dashed) and tensor (solid-gray) modes. Results are shown for the $s=0.3$ case.}
\end{figure}

The form of the power spectrum contribution is further studied for the $s\!=\!0.3$ case in 
Fig.\ \ref{fig:TTdecomp}, which includes a scalar-vector-tensor breakdown. This shows that the broad peak at $\ell\!\approx\!150\!-\!400$ stems from both vector and scalar modes, which peak at $\ell\!\approx\!180$ and $\ell\!\approx\!400$ respectively. The power spectrum is hence dominated by vector modes for all but the smallest scales, with the scalar, vector and tensor contributions having the approximate ratios \mbox{$0.5 : 1.0 : 0.2$} at $\ell\!=\!10$.

\begin{figure}
\resizebox{\columnwidth}{!}{\includegraphics{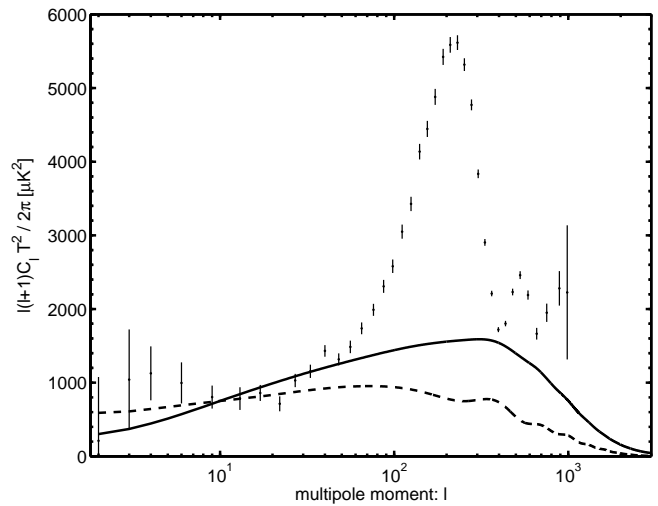}}
\caption{\label{fig:WMAPcomparison}A comparison of the form of the $s=0.3$ string contribution
(solid) to the WMAP results \cite{Hinshaw:2006ia}, enabled by an excessively large normalization to give a match to the data at $\ell=10$. Also shown are results from our global texture simulations (dashed). Note that the binned WMAP 3-year data are plotted which does not include output for $\ell=10$ and that $T$ is the mean CMB temperature .}
\end{figure}

Although CMB data shows the actual power spectrum to peak at around $\ell\!\approx\!200$, with in fact a local minimum at $\ell\!\approx\!400$ \cite{Hinshaw:2003ex, Hinshaw:2006ia} so that the string contribution must be sub-dominant, a useful comparison of forms is provided by setting the normalization of our results to match the WMAP 3-year data at $\ell=10$, as in Fig.\ \ref{fig:WMAPcomparison}. Also included in this figure are the present global texture results, which match those from independent simulations \cite{Bevis:2004wk, Durrer:1998rw, Pen:1997ae} and form a useful check of our algorithms, as well as providing a case for comparison. The slower decay of local strings within the horizon and the greater importance of their (scalar) UETCs in the radiation era yields a significant change in bias between high and low multipoles, with strings remaining important on smaller scales. The string contribution peaks when the data is close to a minimum and is also most precise. This may suggest that the fractional contribution from these string results would be more tightly constrained that those from textures \cite{Bevis:2004wk}. However, it is also true that the overall form of the string contribution matches the data more closely, a fact that may over-turn such an argument.

The normalization to the WMAP data at the conventional $\ell=10$ (and the COBE mean \mbox{temperature \cite{Mather:1998gm})} gives a value of $2\pi G \VEV^{2}$ as 
\mbox{$[2.04\pm0.06\textrm{(stat.)}\pm0.12\textrm{(sys.)}] \times 10^{-6}$}, which is equal to $G\mu$ \cite{Bogomolnyi:1976}. Given the sub-dominance of the string contribution, a value this large is actually ruled out by the data. Nevertheless, this number is useful for the comparison with results from alternative calculations, presented in the next section. Further, the result is not especially sensitive to the power spectrum normalization, varying merely as its square-root. Given the relative similarity in form with the texture result, then the power spectrum normalization needs to be reduced using a factor \mbox{$\approx0.13$} in order to fall in line with the first-year WMAP data incorporated in the likelihood analysis of Bevis et al. \cite{Bevis:2004wk}. However, this corresponds to a reduction of $G \mu$ using a factor of merely \mbox{$\approx0.4$} and in the absence of a full likelihood analysis with the present string results (see \cite{Bevis:2007gh}), the WMAP-normalization value hence serves as guide-line upper-limit. 

\section{Discussion and Conclusions}

\subsection{Comparison with previous string CMB results}

The key result presented is the form of the string contribution to the CMB temperature power spectrum, as shown in Fig.\ \ref{fig:WMAPcomparison} with WMAP-normalization for comparative purposes. Qualitatively, this form is similar to previous string determinations, including those from Nambu-Goto simulations \cite{Contaldi:1998mx} and the latest results from  unconnected segment models \cite{Pogosian:2006hg}. There is agreement that the basic form of the power spectrum has a roughly constant slope at low multipoles, rising up to a single peak, with subsequent decay at small scales. However, the results from the present field simulations give a broader peak due to a greater relative contribution at large scales. While all three approaches yield results that peak at $\ell\!\approx\!400$, the idealized string methods each give 75\% of this peak value at $\ell\approx\!100$ whereas the present technique yields 75\% of the peak value at a much lower multipole: $\ell\approx\!30$. A comparison of the scalar-vector-tensor sub-contributions relative to the unconnected segment results \cite{Pogosian:2006hg}, highlights that this dissimilarity stems largely from the difference in form of the vector contribution. In the present case the vector component involves a very wide peak, quite different to that seen with the unconnected segment model.

The \mbox{$\ell=10$} WMAP-normalization of the power spectrum result giving $G\mu=(2.04\pm0.13) \times 10^{-6}$ is larger than FRW Nambu-Goto simulation results at large angular scales, for which COBE-normalization gave: \mbox{$G\mu\!=\!(0.7\pm0.2)\times 10^{-6}$ \cite{Landriau:2003xf}} and \mbox{$G\mu\!=\!(1.05^{+0.35}_{-0.20})\times 10^{-6}$ \cite{Allen:1996wi}} (with the dependence of this value upon the cosmology explored in the former). The UETC results of Contaldi et al. \cite{Contaldi:1998mx}, involving Minkwoski space-time appears to be consistent with these, being \mbox{$G\mu\!=\!1.0\times 10^{-6}$} ($\ell\!=\!5$). Finally this is true also for the latest unconnected segment result of \mbox{$G\mu\!=\!1.1\times 10^{-6}$ \cite{Pogosian:2006hg}} (normalized to the total WMAP power but, by chance, an approximately equivalent normalization).

It is difficult to be certain as to the cause of the differences in form between the field evolution and idealized string results but possibilities include the treatment of the decay products or differences in velocity correlations. Although we consider the former to be more likely, it is interesting to note that a recent study of Nambu-Goto simulations in both Minkowski and FRW metrics \cite{Martins:2005es} has highlighted a difference with regard to velocity correlations and it is true that unconnected segment models cannot model such correlations. We hence advise some caution with regard to the interpretation of such results, which do not include the same level of physics as incorporated in the present simulations. However, the extrapolation involved here via scaling is over many orders of magnitude. Although it must be employed in all such calculations covering the full range of CMB scales and finds justification from the simulations themselves, its requirement represents an inability to probe all scales involved to the desired degree.

\subsection{Comparison with global strings}

Global strings have previously been shown to give a contribution that is similar to that from global textures, albeit that the result peaks at slightly larger multipoles \cite{Pen:1997ae}. 
Fig.\ \ref{fig:WMAPcomparison} then indicates that global strings are intermediate in form between textures and the present local string results, but are actually somewhat closer to the former. Global strings have a less localized energy distribution than gauge strings and so experience significant long range forces \cite{Perivolaropoulos:1991du}. This presumably causes the global case to show a more rapid decay within the horizon and their CMB contribution to have less significance at small scales, as seen in such calculations \cite{Pen:1997ae}.

\subsection{Future prospects}

It is not just the temperature power spectrum that is important. Significant constraints on cosmic string scenarios might some day arise from the measures of the polarization of CMB photons. Particularly, the so-called B-mode polarization spectrum provides an important window on cosmic strings because inflation contributes to this only weakly. Scalar modes may contribute to the B-mode only via the gravitational lensing of the E-mode signal, with a second inflationary contribution arising from the sub-dominant tensor modes. It may hence be the case that the large vector contributions from cosmic strings enable their signature to be detected using data from future B-mode projects \cite{Taylor:2004hh, Planck, Polarbear} and this makes the difference in vector mode results especially interesting. Presently, B-mode results have only been published for unconnected segment models \cite{Pogosian:2006hg}, with no direct input yet from Nambu-Goto simulations and therefore the situation is a little different in the polarization case. We will present polarization results from field simulations in a forthcoming publication \cite{Bevis:2006b}. 

The future also holds a great deal for the temperature power spectrum with, for example, the planned full-sky coverage at sub-WMAP scales from the Planck satellite \cite{Planck}. Such data  will help to restrict the inflationary contribution to the CMB and so more heavily constrain any sub-dominant component from cosmic strings. However, it should be noted that CMB perturbations from cosmic strings are non-Gaussian and are not statistically summarized using the power spectrum alone (differing in that respect from inflationary models under linear perturbation theory). Hence it is also important to consider predictions for cosmic strings beyond the power spectrum, which our present UETC approach is unable to provide, but that may enable additional constraints.


\begin{acknowledgments}
We acknowledge support from PPARC (N.B., M.H, M.K.), the National Science Foundation, USA (J.U.), and the ESF COSLAB programme, FPA2005-04823 and 9/UPV00172.310-14497/2002. Production simulations were run on the UK National Cosmology Supercomputer, supported by SGI, Intel, HEFCE and PPARC. We would like to thank Stuart Rankin and Victor TraviesoVictor Travieso, who run that facility, and acknowledge useful discussions with Richard Battye, James Karamath, Andrew Liddle, Carlos Martins and Paul Shellard.
\end{acknowledgments}


\appendix

\section{Systematic error analysis}

In this appendix we detail the numerical experiments on which we base our error estimates, summarized in Sec.\ \ref{sec:CMBPowSpe}. The experiments are compared to our production runs, in which the parameters chosen are laid out in Table \ref{tab:ProdParms}. 

\subsection{Conformal time offset $\tau_{\xi=0}$}

The scaling form of the UETC functions (see Eqs. \ref{eqn:C11} - \ref{eqn:CT}) is extracted from the energy momentum correlators through multiplication by powers of the shifted conformal time $\tau - \tau_{\xi = 0}$, as described around Eq.\ \ref{eqn:XiOffset}.  Errors in the determination of the offset time $\tau_{\xi = 0}$ will feed through into errors in the scaling functions.  For the primary results, the offset time is found via a linear fit of the string length parameter $\xi(\tau)$ over the range $64\VEV^{-1} < \tau_{\mathrm{sim}} < 128\VEV^{-1}$. We estimate the errors by fitting over narrower ranges of conformal time, as shown in Table \ref{tab:XiFit}.  These errors can also be viewed as an estimate of the errors from not reaching true scaling as $\tau \to \infty$.

\subsection{Dynamic range}

Our data is taken over quite a limited dynamic range, defined as the maximum ratio $R = \tau/\tau'$, where $\tau$ and $\tau'$ are the two times in the UETC. Before the offset is taken into account, this ratio is 2.5, but is reduced to about 1.8 with the offset (and is slightly different for each run).  In order to estimate the errors from the limited dynamic range we further truncate the UETCs at smaller values of $R$, with results listed in Table \ref{tab:rRange}.  We immediately see that the power spectrum increases with decreasing $R_{\mathrm{max}}$,  and that there is evidence that $R_{\mathrm{max}} \approx 1.8$ overestimates the power. A logarithmic fit of the error against $R_{\mathrm{max}}$ indicates that, at least for $\ell = 10$ and $\ell =100$, the power spectrum is converging to a value about 10\% lower than at $R_{\mathrm{max}} \approx 1.8$. This is our estimate of the error due to the dynamic range.  It is interesting that truncating the UETCs \emph{increases} the power: this may well be due to the truncation producing ``ringing'' in Fourier space, which artificially sources extra perturbations.

\begin{table}
\begin{ruledtabular}
\begin{tabular}{lc}
Parameter		                	                             &	Value\\
\hline
Lattice size $N$		                      	               &	512\\
Lattice spacing	$\Delta x$ [$\VEV^{-1}$]	                 &	0.5\\
Timestep	$\Delta \tau$ [$\VEV^{-1}$]			                 &	0.1\\
Scalar coupling $\lambda_0$		                             &	2.0\\
Gauge coupling $e_0$		                                   &	1.0\\
Initial conformal time	 $\tau_{\mathrm{i}}$ [$\VEV^{-1}$] &	1.0\\
Final conformal time $\tau_{\mathrm{e}}$ [$\VEV^{-1}$]     &	160\\
Initial $s$					                                       &	-0.116\\
Final $s$					                                         &	0.3\\
Time of change in $s$ value [$\VEV^{-1}$]	                 &	32\\
&\\
$\tau_{\mathrm{sim}}$ range of $\xi$ fit	[$\VEV^{-1}$]	   &	64 - 128\\
Dynamic range $R_{\rm max} = (\tau/\tau^{\prime})_{\mathrm{max}}$ &	$\approx 1.8$\\
Eigenvector decomposition matrix size                      &	512\\
Eigenvector decomposition matrix $k\tau$ spacing           &	linear\\
No. terms used in eigen-contribution sum                   &	128\\
\end{tabular}
\end{ruledtabular}
\vspace{-0.2cm}
\caption{\label{tab:ProdParms} Parameters used in production runs.  The first part of the table lists the parameters of the Abelian Higgs model simulations, described in Sec.\ \ref{sec:simulations}, with the 
second part listing parameters of the UETC method for calculating the CMB power spectrum as 
described in Secs.\ \ref{sec:results}, \ref{sec:CMBPowSpe}.}
\par
\vspace{0.5cm}
\begin{ruledtabular}
\begin{tabular}{c|cccc}
$\tau$ range & $\ell=10$ & $\ell=100$ & $\ell=300$ & $\ell = 1000$ \\
  \hline
64 - 128		&	 0   &	0   &	0   &  0\\
64 - 112		&	-6.9&	-7.8&	-6.7& -5.7\\
64 - 96			&	-8.7&	-9.4&	-8.6& -9.7\\
\end{tabular}
\end{ruledtabular}
\vspace{-0.2cm}
\caption{\label{tab:XiFit} The percentage change in the temperature power spectrum when changing the conformal time range over which the string length parameter $\xi$ is fitted.}
\par
\vspace{0.5cm}
\begin{ruledtabular}
\begin{tabular}{c|cccc}
$R_{\rm max}$ & $\ell=10$ & $\ell=100$ & $\ell=300$ & $\ell = 1000$ \\
  \hline
$\approx 1.8$	&0	&0	&0	&0	\\
\ \ \ 1.6	&7.0	&5.5	&4.8	&8.1	\\
\ \ \ 1.5	&13	&10	&5.9	&11	\\
\ \ \ 1.4	&22	&16	&9.8	&14	\\
\end{tabular}
\end{ruledtabular}
\vspace{-0.2cm}
\caption{\label{tab:rRange} The percentage change in the temperature power spectrum when changing the
dynamic range parameter $R_{\mathrm{max}} = (\tau/\tau')_{\mathrm{max}}$.}
\par
\vspace{0.5cm}
\begin{ruledtabular}
\begin{tabular}{c|cccc}
$s$ & $\ell=10$ & $\ell=100$ & $\ell=300$ & $\ell = 1000$ \\
  \hline
0.3	&0	&0	&0	&0	\\
0.2	&13	&8.2	&0.7	&-7.2	\\
0.0	&2.6	&-1.1	&-2.0	&1.1	\\
\end{tabular}
\end{ruledtabular}
\vspace{-0.2cm}
\caption{\label{tab:Fat} The percentage change in the temperature power spectrum when changing the
string width shrinkage parameter $s$.}
\par
\vspace{0.5cm}
\begin{ruledtabular}
\begin{tabular}{c|cccc}
 & $\ell=10$ & $\ell=100$ & $\ell=300$ & $\ell = 1000$ \\
  \hline
Matter \& radiation &0	&0	&0	&0	\\  
Matter only &0.03	&-3.8	&-11	&-14	\\  
\end{tabular}
\end{ruledtabular}
\vspace{-0.2cm}
\caption{\label{tab:matterrad} The percentage change in the temperature power spectrum when using 
matter era eigenvectors throughout the simulation, instead of interpolating to radiation 
era at $\tau < \tau_{\rm eq}$.}
\end{table}

\subsection{String width modification}

In order to deal with the shrinking of the string width in comoving units, we modify the equations of motion so that the strings grow in physical units. This growth is parametrized by $s$, the power of the scale factor by which the comoving width shrinks (Eqs.\ \ref{eqn:AHdynamicPhiS}, \ref{eqn:AHdynamicFS}). As seen in table V and also in Fig. \ref{fig:ETCvaryS}, there is no obvious trend to extrapolate and, as discussed in detail in Sec. V, the systematic effect is likely to be comparable to the statistical uncertainties shown in the figure.

\subsection{Matter-radiation transition}

It is not obvious in the UETC method what to do when the expansion rate changes.  We interpolate between radiation era and matter era eigenfunctions.  In order to estimate the error associated with this procedure we calculated the power spectrum with matter era eigenfunctions only, which had the effect of decreasing the power by up to 14\% at $\ell=1000$, but was negligible at $\ell=10$ (see table \ref{tab:matterrad}).
\begin{table}
\begin{ruledtabular}
\begin{tabular}{c|cccc}
No.\ eigenvectors & $\ell=10$ & $\ell=100$ & $\ell=300$ & $\ell = 1000$ \\
  \hline
128	&0	&0	&0	&0	\\  
150	&0.04	&0.09	&0.1	&0.2	\\
\end{tabular}
\end{ruledtabular}
\vspace{-0.2cm}
\caption{\label{tab:EigSum} The percentage change in the temperature power spectrum when changing 
the number of eigenvectors in the sum.}
\par
\vspace{0.5cm}
\begin{ruledtabular}
\begin{tabular}{c|cccc}
Size of matrix & $\ell=10$ & $\ell=100$ & $\ell=300$ & $\ell = 1000$ \\
  \hline
512	&0	&0	&0	&0	\\  
256	&0.4	&0.2	&-0.4	&0.2	\\
128	&15	&-1.9	&-8.8	&-16	\\
64	&1.6	&-16	&-24	&-33	\\
\end{tabular}
\end{ruledtabular}
\vspace{-0.2cm}
\caption{\label{tab:MatrixSample} The percentage change in the temperature power spectrum when changing 
eigenvector decomposition matrix size $M$.}
\par
\vspace{0.5cm}
\begin{ruledtabular}
\begin{tabular}{c|cccc}
Type (No.\ e-vectors) & $\ell=10$ & $\ell=100$ & $\ell=300$ & $\ell = 1000$ \\
  \hline
Linear (128)	& 0 & 0 & 0 & 0 \\
Log (128)	&-1.7	&-5.3	&-7.3	&-23	\\  
Log (400)	&0.3	&0.8	&-0.8	&-11	\\  
\end{tabular}
\end{ruledtabular}
\vspace{-0.2cm}
\caption{\label{tab:MatrixSampling} The percentage change in the temperature power spectrum when changing 
eigenvector decomposition matrix sampling from linear to logarithmic spacing in $k\tau$.}
\par
\vspace{0.5cm}
\begin{ruledtabular}
\begin{tabular}{lr}
 Source &Value \\
  \hline
Conformal time offset	& $+9$\% \\
Dynamic range			& $-10$\%\\
String width modification	& $\pm5$\%\\
Matter-radiation		& $\pm11$\%\\
Eigenvector sum		& $<1$\%\\
Eigenvector decomposition matrix size		& $<1$\%\\
Eigenvector decomposition matrix $k\tau$ spacing	& $<1$\%\\
&\\
Total in quadrature		&  $\pm16$\%\\
\end{tabular}
\end{ruledtabular}
\vspace{-0.2cm}
\caption{\label{tab:SysErrs} Estimates of systematic errors in the temperature power spectrum at $\ell=300$
due to the sources discussed above.}
\end{table}

\subsection{Eigen-contribution sum convergence}

Fig.\ \ref{fig:TTconverge} shows how the power spectrum changes as increasing numbers of eigenvectors of the UETC matrices are included. To estimate how close we are to convergence we increased the number of eigenvectors from 128 to 150, finding the negligible changes given in Table \ref{tab:EigSum}.

\subsection{Eigenvector decomposition matrix size and spacing}

The UETCs are recorded at discrete values of $k\tau$ and $r=\tau/\tau'$, but must be diagonalized as a matrix with rows and columns labeled by $k\tau$ and $k\tau'$. The number and spacing of the $k\tau$ (and $k\tau'$) values used to construct this matrix are not fundamental and must be chosen to give reasonably accurate results while requiring a minimal number of eigen-contributions for the convergence of the power spectrum. It seems that the size of the matrix barely affects the power spectrum above a value of 256, as can be seen from Table \ref{tab:MatrixSample}.  We tried linear and logarithmic sampling, finding that there was negligible difference providing enough eigenvalues were taken in the sum, as shown in Table \ref{tab:MatrixSampling}. In view of the smaller number of eigenvalues linear sampling is preferred, and we conclude that negligible errors are associated with the sampling.

\subsection{Summary}

The important range for future fitting to WMAP data will be $\ell\approx300$, for which we
summarize the errors in Table \ref{tab:SysErrs}. Note that the conformal time offset error gives an underestimate while the dynamic range error means that our results are likely to be an overestimate. We obtain final estimates for the upward and downward errors by combining these with the other errors in quadrature. When applied to the WMAP normalization value of $G\mu$ we find $G\mu = [2.04\pm0.06(\mathrm{stat.})\pm0.12(\mathrm{sys.})]\times10^{-6}$, recalling that the temperature power spectrum depends on the square of $G\mu$.

\bibliography{references}

\end{document}